\documentclass[%
 reprint,
 amsmath,amssymb,
 aps,
]{revtex4-2}

\usepackage{graphicx}
\usepackage{dcolumn}
\usepackage{bm}
\usepackage{hyperref}
\usepackage{multirow}
\usepackage{blkarray}
\usepackage{ulem}

\usepackage{tikz}
\usetikzlibrary{patterns}
\usetikzlibrary{calc}
\usetikzlibrary{decorations.pathreplacing,decorations.markings}
\usetikzlibrary{arrows}

\tikzset{
  on each segment/.style={
    decorate,
    decoration={
      show path construction,
      moveto code={},
      lineto code={
        \path [#1]
        (\tikzinputsegmentfirst) -- (\tikzinputsegmentlast);
      },
      curveto code={
        \path [#1] (\tikzinputsegmentfirst)
        .. controls
        (\tikzinputsegmentsupporta) and (\tikzinputsegmentsupportb)
        ..
        (\tikzinputsegmentlast);
      },
      closepath code={
        \path [#1]
        (\tikzinputsegmentfirst) -- (\tikzinputsegmentlast);
      },
    },
  },
  mid arrow/.style={postaction={decorate,decoration={
        markings,
        mark=at position .65 with {\arrow[#1]{stealth}}
      }}},
}

\newcommand{\p}{\partial}
\newcommand{\vx}{\mathbf{x}}
\newcommand{\vxp}{\vx'}
\newcommand{\va}{\mathbf{a}}
\newcommand{\vb}{\mathbf{b}}
\newcommand{\vc}{\mathbf{c}}
\newcommand{\vq}{\mathbf{q}}
\newcommand{\vQ}{\mathbf{Q}}
\newcommand{\vp}{\mathbf{p}}
\newcommand{\vv}{\mathbf{v}}
\newcommand{\vu}{\mathbf{u}}
\newcommand{\vf}{\mathbf{f}}
\newcommand{\vj}{\mathbf{j}}
\newcommand{\vL}{\mathbf{L}}
\newcommand{\vom}{\mathbf{w}}
\newcommand{\vnabla}{\mathbf{\nabla}}
\newcommand{\tvom}{\bar{\mathbf{w}}}
\newcommand{\tvv}{\bar{\mathbf{v}}}
\newcommand{\tvu}{\bar{\mathbf{u}}}
\newcommand{\tvj}{\bar{\mathbf{j}}}
\newcommand{\tvL}{\bar{\mathbf{L}}}
\newcommand{\ttheta}{\bar{\theta}}
\newcommand{\tv}{\bar{v}}
\newcommand{\tu}{\bar{u}}
\newcommand{\tj}{\bar{j}}
\newcommand{\tK}{\bar{K}}
\newcommand{\tom}{\bar{w}}
\newcommand{\bla}{\big\langle}
\newcommand{\bra}{\big\rangle}
\newcommand{\la}{\langle}
\newcommand{\ra}{\rangle}
\newcommand{\nd}{\noindent}

\begin{document}

\title{The inviscid fixed point of the multi-dimensional Burgers-KPZ equation}

\author{Liubov Gosteva$^1$, Malo Tarpin$^2$, Nicol\'as Wschebor$^3$, L\'eonie Canet$^{1,4}$}
\affiliation{$^1$ Universit\'e Grenoble Alpes, CNRS, LPMMC, 38000 Grenoble, France\\ $^2$ Ecole Centrale de Lyon, CNRS, Universite Claude Bernard Lyon 1, INSA Lyon, LMFA, 69130, Ecully, France\\ $^3$ Instituto de F\'isica, Facultad de Ingenier\'ia, Universidad de la Rep\'ublica, J.H.y Reissig 565, 11000 Montevideo, Uruguay\\ $^4$ Institut Universitaire de France, 5 rue Descartes, 75005 Paris}


\begin{abstract}

A new scaling regime characterized by a $z=1$ dynamical critical exponent has been reported in several numerical simulations of the one-dimensional Kardar-Parisi-Zhang and noisy Burgers equations. In these works, this scaling, differing from the well-known KPZ one $z=3/2$, was found to emerge in the tensionless limit for the interface and in the inviscid limit for the fluid. Based on functional renormalization group, the origin of this scaling has been elucidated. It was shown to be controlled  by a yet unpredicted fixed point of the one-dimensional Burgers-KPZ equation, termed inviscid Burgers (IB) fixed point. The associated universal properties, including the scaling function, were calculated. All these findings were restricted to  $d=1$, and it raises the intriguing question of the fate of this new IB fixed point in higher dimensions. In this work,  we address this issue and analyze the multi-dimensional Burgers-KPZ equation using functional renormalization group. We show that the IB fixed point exists in all dimensions $d\geq 0$, and that it controls the large momentum behavior of the correlation functions in the inviscid limit. It turns out that it yields in all $d$  the same super-universal value $z=1$ for the dynamical exponent.
 \end{abstract}

\maketitle

\section{Introduction}

The Kardar-Parisi-Zhang (KPZ) universality class is impressive by its broadness, which keeps extending. Beyond the realm of stochastically growing interfaces -- for which the KPZ equation was originally derived~\cite{Kardar86} -- the KPZ scaling has been both unveiled theoretically and observed experimentally in many quantum systems ranging from integrable systems, such as Heisenberg quantum spin chains~\cite{Bloch2022,Tennant2022}, to driven-dissipative systems, such as exciton-polariton condensates~\cite{Fontaine2022Nat}. It has also emerged in quantum information in the scaling of entanglement growth~\cite{Nahum2017} or in Anderson localization in the scaling of wave-packets density fluctuations~\cite{Mu2024}.
 This broadness was first recognized for classical systems~\cite{Halpin-Healy95,Krug97}, since the KPZ equation  maps to
 the noisy Burgers equation for randomly stirred fluid~\cite{Burgers48}, as well as to the model of  directed polymers in random media~\cite{Kardar87}. All of these problems are whole research fields in themselves, and a lot of efforts have been devoted to understanding the rich physics of each.

The hallmark of the KPZ universality class in dimension $d=1$ is the superdiffusive scaling $z=3/2$~\cite{Kardar86}, which  was first reported for the noisy Burgers equation in Ref.~\cite{Forster76,Forster77}, where $z$ is the dynamical critical exponent, defined as the anomalous scaling between space and time  $t\sim x^z$.
 This scaling $z=3/2$ differs from the diffusive equilibrium one $z=2$, which  corresponds to the scaling of the linear Edwards-Wilkinson (EW) equation~\cite{Edwards82}.
In the KPZ equation, the source of noise is microscopic, in the sense that it predominantly acts at small scales. In this case,  one is interested in the large-distance long-time behavior of the interface.
 In contrast, in the context of the Burgers equation, one rather considers a large-scale stochastic forcing, and the regime of interest is the inertial range, which corresponds to the small-scale short-time behavior. In this regime, the system is characterized by a
 kinetic energy spectrum $E(k)\sim k^{-2}$, and strong intermittency effects~\cite{Verma2000}.
  Intermittency effects are usually evidenced through static equal-time  statistics. In particular, a prominent example are the structure functions,  whose scaling reflects a characteristic bi-fractal distribution of anomalous exponents~\cite{Bec2007}.

  The temporal properties are much less discussed.
 In this respect, a pioneering work was carried out in Ref.~\cite{Bouchaud1995} using the mapping from the noisy Burgers equation to the disordered problem of directed polymers, and applying the replica method to compute averages over the disorder. It led to a solution for the correlation function obtained from a variational ansatz  featuring a $z=1$ dynamical exponent, which was interpreted as a convective scaling
  associated with the formation of celular patterns. This variational solution is expected to become exact in the limit of infinite dimension $d\to\infty$ and infinite Reynolds number (inviscid limit).  Whereas this result was obtained in the case of  a large-scale forcing, a similar $z=1$ dynamical exponent was also uncovered  in different simulations of Burgers~\cite{Brachet2022,Majda2000} or KPZ~\cite{Rodriguez2022} equations  with short-scale microscopic noise in $d=1$ and in the inviscid (respectively, tensionless) limit. Moreover, the same scaling was found in a strongly interacting one-dimensional quantum bosonic system belonging to the KPZ class~\cite{Fujimoto2020}. Using functional renormalization group (FRG)~\cite{Fontaine2023InvBurgers}, this scaling $z=1$ was shown to result from the existence of a yet unpredicted fixed point of the KPZ equation, termed the inviscid Burgers (IB) fixed point. The latter FRG analysis on the IB fixed point was focused on $d=1$. This raises the natural question of what happens to this new fixed point in higher dimensions, and whether it still coexists with the KPZ fixed point. Another intriguing issue is whether there is a connection between the $z=1$ scaling observed in $d=1$ and the one reported in $d\to \infty$.

In this work, we answer  these questions. We show that the IB fixed point exists in all dimensions, and  independently of the scale of the forcing, as long as  the latter remains Gaussian. This bridges the result of~\cite{Bouchaud1995}
 valid for $d\to\infty$ and large-scale forcing with the ones of~\cite{Brachet2022,Rodriguez2022,Fujimoto2020,Fontaine2023InvBurgers} obtained for $d=1$ and small-scale noise.
 This fixed point controls the large momentum behavior (UV scales, short distances) of the correlation functions in the limit $\nu\to0$, and it yields a $z=1$ dynamical exponent in all dimensions.

 In fact, a similar $z=1$ scaling was also unveiled in the large-momentum behavior of correlation functions in Navier-Stokes turbulence~\cite{Canet2017,Tarpin2018}, and was related in this context to the random sweeping effect, that is, the sweeping of small-scale velocities by the large-scale eddies of the turbulent flow. We show in this paper that this common scaling is ultimately rooted in the Galilean invariance and the existence of a non-Gaussian fixed point. Our derivation of the $z=1$ dynamical exponent in the multi-dimensional Burgers-KPZ equation stems from FRG, and deeply exploits the Ward identities associated with the symmetries of the underlying dynamics, in a similar way as for the Navier-Stokes problem.
 
The remaining of the paper is organized as follows. We construct in Sec.~\ref{sec:action} an appropriate action for the $d$-dimensional Burgers equation, and discuss its symmetries and the associated Ward identities in Sec.~\ref{sec:Ward}. We introduce in  Sec.~\ref{sec:FRG} the FRG formalism, and elucidate in  Sec.~\ref{sec:LPA}, based on a simple approximation, the fixed-point structure of the Burgers-KPZ equation, showing in particular the existence of the IB fixed point in all $d$. In  Sec.~\ref{sec:largep}, we show that the flow equation for the two-point velocity correlation function can be closed in the large momentum $p$ limit, and this closure is asymptotically exact for $p\to\infty$. We finally present the fixed-point solution of this flow equation and show that it features the $z=1$ dynamical exponent in all $d$.

\section{Field theory formulation of the stochastic Burgers-KPZ equation}

\subsection{Action for the $d$-dimensional Burgers equation}
\label{sec:action}

The stochastic Burgers equation~\cite{Burgers48} is a simplified version of the Navier-Stokes equations for a potential flow -- which satisfies $\vnabla \times \vv = 0$ -- without pressure terms, and stirred by a stochastic force $\vf$. It reads as
\begin{equation}\label{eq:stochEq}
    \partial_t \vv + \vv \cdot \vnabla \vv  = \nu \nabla^2 \vv + \vf\,,
\end{equation}
with $\nu$ the kinematic viscosity of the fluid.
The random force is chosen to be Gaussian distributed with zero mean and  covariance
\begin{equation}
\label{eq:forcing}
\bla f_\alpha(t,\vx) f_\beta(t',\vxp)\bra  = 2  \delta(t-t')N_{\alpha\beta}(\vx-\vxp)\,,
\end{equation}
where $N$ is centered around a large scale $L$ to generate turbulence.
Since the velocity is irrotational, one can define a scalar function $h$ such that $\vv = - \lambda \vnabla h$ with $\lambda$ a real parameter. The dynamics of this field, interpreted as the height of an interface, is then given by the KPZ equation
\begin{equation}
\label{eq:KPZ}
    \partial_t h = \nu \nabla^2 h + \frac{\lambda}{2}(\nabla h)^2 + \eta\,.
\end{equation}
In the original KPZ equation~\cite{Kardar86},  $\eta$ is a Gaussian noise of zero mean and covariance
\begin{equation}
\label{eq:etaKPZ}
\bla \eta(t,\vx) \eta(t',\vxp)\bra = 2 D\delta(t-t')\delta^d(\vx-\vxp)\,.
\end{equation} 
Through the mapping to the Burgers equation, one obtains that the stochastic forcing is related to the KPZ noise as $\vf = -\lambda\vnabla\eta$. Thus,  the covariance \eqref{eq:etaKPZ}  corresponds for the Burgers equation to $\vf$ being akin a thermal noise  -- implying in particular that it predominantly acts at small scales.

In order to apply  renormalization group techniques, one can cast the stochastic equation into a path integral formulation which encompasses all the trajectories emanating  from different noise realizations, following the  Martin-Siggia-Rose-Janssen-de Dominicis  formalism~\cite{Martin73,Janssen76,Dominicis76}.
For the KPZ equation, this is straightforward and one obtains~\cite{Frey94}
\begin{align}
{\cal Z}[j,\tj] &= \int {\rm D} h {\rm D} \bar{h} e^{-{\cal S}_{\rm \tiny KPZ}[h,\bar{h}] + \int_{t,\vx}\{ j h + \tj \bar{h}\}}\, ,\nonumber\\
{\cal S}_{\rm \tiny KPZ}[h,\bar{h}] &= \int_{t,\vx}\Big\{\bar{h} \Big[\p_t h -\frac \lambda 2 (\nabla h)^2 - \nu \nabla^2 h   \Big] - D\,\bar{h}^2\Big\}\, .
\label{eq:actionKPZ}
\end{align}
For the Burgers equation, the construction of an action is more subtle and is addressed in the following. The two formulations are striclty equivalent in principle. However, although the KPZ action is much simpler, it turns out that the exact closure of the FRG equations at large momentum (presented in Sec.~\ref{sec:largep}) is much easier in the Burgers formulation. The reason is that one of the symmetries (the time-gauged shift of the response fields discussed in Sec.~\ref{sec:Ward}) needed to complete this closure is transparent in the Burgers formulation, whereas it is somehow hidden in the KPZ one. We do not understand at the moment how to achieve the exact closure in the KPZ formulation. We thus choose to work with the Burgers action in the rest of the paper for consistency (eventhough within the simple approximation presented in Sec.~\ref{sec:LPA}, calculations can also be done directly in the KPZ formulation \cite{Canet2005b}).

We now present how to build an action for the Burgers equation. In $d=1$, it is straightforward, following the same  Martin-Siggia-Rose-Janssen-de Dominicis  procedure, and one obtains the action~\cite{Fontaine2023InvBurgers}
\begin{equation}
{\cal S}[v,\tv] = \int_{t,x}\Big\{\tv \Big[\p_t v + v \p_x v - \nu \p^2_x v   \Big] - {\cal D}\big(\p_x\tv \big)^2\Big\}\, ,
\label{eq:actionBurgers1D}
\end{equation}
where ${\cal D} = D \lambda^2$, and the conservative form of the noise term is inherited from the mapping with the KPZ equation.
In higher dimension $d$, one should further impose the irrotationality constraint.
In the seminal work of Forster-Nelson-Stephen~\cite{Forster77}, the one-dimensional Burgers equation is formally continued to arbitrary dimensions by using the identity $\vv\cdot\vnabla \vv = \frac 1 2 \vnabla \vv^2 - \vv\times (\vnabla\times \vv)$ and deleting the last term. However, in practice, only the one-dimensional case is studied  in Ref.~\cite{Forster77}. Since the FRG calculations presented in the following involve arbitrary $n$-point correlation functions, the irrotationality constraint must be explicitly encoded in the action. This can be simply achieved by introducing another response field $\tvom$ which acts as a Lagrange multiplier for this constraint. This procedure is similar to the one used to impose incompressibility in the Navier-Stokes action~\cite{Canet2016,Canet2022}.
Furthermore, one can check from the KPZ mapping that the response velocity field $\tvv$ should also be irrotational. We hence introduce another  response field $\vom$ to impose this constraint.
However, it turns out that the action resulting from adding these two auxiliary Lagrange multipliers is ill-defined as its Hessian is not invertible, which implies that the propagator does not exist, preventing any field-theoretical calculation. This is due to the presence of spurious zero modes, which originate in the fact that the $\tvom$ and $\vom$ fields are coupled only through their transverse parts. To fix their longitudinal parts and thereby remove the zero modes, one can introduce the additional scalar Lagrange multiplier fields $\theta$, $\ttheta$. Thus, we obtain the following action for the Burgers-KPZ equation
\begin{align}
&{\cal Z}[{\cal J}] = \int {\rm D} \Phi\; e^{-{\cal S}[\Phi] + \int_{t,\vx} {\cal J}\Phi }\label{eq:Z}\\
    \mathcal{S}[\Phi] &= \int_{t,\vx} \Big\{
        \tvv \cdot \left[
            \partial_t \vv + (\vv \cdot \nabla) \vv - \nu \nabla^2 \vv
        \right]
        - {\cal D} (\nabla \cdot \tvv)^2 \nonumber\\
      &  + \tvv \cdot (\nabla \times \vom)  + \tvom \cdot (\nabla \times \vv) +  \ttheta \nabla \cdot \vom
        + \tvom \cdot \nabla \theta \Big\}\, ,\label{eq:action}
\end{align}
(`-KPZ' referring to the fact that  the conservative form of the forcing inherited from the mapping to KPZ is kept).
 Here and in the following,  $\Phi$ denotes the multiplet of fields $(\vv,\tvv,\vom,\tvom,\theta,\ttheta)$, ${\cal J}$ the multiplet of associated sources $(\vj,\tvj,\vL,\tvL,K,\tK)$, and summation over all fields is implicit in the notation ${\cal J}\Phi$. Note that, in the presence of a large-scale forcing \eqref{eq:forcing} instead of the KPZ-inherited noise term, the latter can be simply replaced with $\int_{t,\vx,\vx'} \tv_\alpha(t,\vx)N_{\alpha\beta}(\vx-\vxp)\tv_\beta(t,\vx')$, which then provides the general Burgers action.

Let us note that we have conveniently used vectorial product and mixed product notations, which are appropriate in $d=3$ (or $d=2$). However, this can be  extended to arbitrary dimensions by expressing any product of the form $\vc\cdot (\va\times\vb)$ in an anti-symmetrized form, such as $c_{\alpha\beta}(a_\alpha b_\beta-b_\alpha a_\beta)$ where $c$ is now a rank-two tensor. For dimensions higher than $d=3$, it requires to introduce supplementary constraints to keep the Hessian of the action invertible. We show in Appendix~\ref{app:higherd} how this can be achieved without changing the discussion below. For simplicity, we keep using the notation of~(\ref{eq:action}) in the following, and also express
vectorial products using the Levi-Civita tensor $\epsilon_{\alpha\beta\gamma}$, bearing in mind that they can be replaced by anti-symmetric products in arbitrary dimensions $d>1$.
 All the results presented in the following are thus valid in any $d$.

\subsection{Extended symmetries and Ward identities}
\label{sec:Ward}

In order to constrain the FRG approximations, we are interested not only in symmetries in the strict sense, but also in extended symmetries. Extended symmetries
 are related to transformations that do not leave the action  invariant, but induce a variation linear in the ﬁelds.
 These extended symmetries allow one to derive  exact identities on correlation functions, under the form of Ward identities, which are typically  more constraining than their non-extended counter-parts. The Ward identities follow from the requirement that the symmetry transformations do not affect the measure of the path integral (\ref{eq:Z})~\cite{Canet2015,Canet2022}. This leads to  functional identities satisfied by  ${\cal W} = \ln{\cal Z}$, the generating functional of connected correlation functions -- which are  the equivalent,for fluctuating fields instead of random variables, of cumulants. Equivalently, Ward identities can be derived for  the Legendre transform of ${\cal W}$, which is called the effective action and denoted  ${\Gamma}$. The effective action is the generating functional of one-particle irreducible correlation functions, also termed vertex functions. It is defined by
\begin{equation}
    \Gamma[\Psi]  =  \underset{\mathcal{J}}{\sup} \left[
        \int_{t,\vx}
        \mathcal{J} \Psi
        -\mathcal{W}[\mathcal{J}]
    \right]
\end{equation}
where
\begin{equation}
\label{eq:defpsi}
 \Psi_i\equiv\langle\Phi_i\rangle_{\cal J}=\dfrac{\delta {\cal W}}{\delta {\cal J}_i}
\end{equation}
are the expectation values of the fields in the presence of the sources ${\cal J}$.
The knowledge of the set of connected correlation functions is striclty equivalent to the one of  one-particle irreducible correlation functions. However, from a field-theoretical viewpoint, the advantage of working with the latters instead of the formers is that less and simpler diagrams are involved in any  diagrammatic calculations. The FRG is thus naturally   formulated in terms of vertex functions, so that we express in the following the Ward identities on the effective action $\Gamma$ rather than on ${\cal W}$.

The Burgers-KPZ action~(\ref{eq:action}) possesses  three extended symmetries,  in close analogy with the Navier-Stokes action~\cite{Canet2015,Canet2022}, that we now expound.
Note that the form of these extended symmetries is independent of the precise profile of the noise as long as it is Gaussian and white in time. As a consequence, all the Ward identities derived in the following for $n$-point vertices ($n>2$) hold both for a short-scale noise and for a large-scale forcing {\it i.e.} both for the Burgers-KPZ and the Burgers actions.

\subsubsection{Fully-gauged shift symmetries of the auxiliary fields.}

The four transformations
\begin{equation}
\label{eq:fullshift}
    \varphi(t,\vx) \rightarrow
    \varphi(t,\vx) + \varepsilon_{\varphi}(t,\vx)
    , \; \varphi \in
    \{\vom, \tvom, \theta, \ttheta \}\,,
\end{equation}
where $\varepsilon_{\varphi}$ is either a scalar (for $\varphi=\theta, \ttheta$) or a vectorial (for $\varphi=\vom, \tvom$) infinitesimal field, are  extended symmetries of the Burgers action. Indeed, as the terms involving the auxiliary fields are quadratic, their variation under a shift is automatically at most linear.  Since each transformation~\eqref{eq:fullshift} is a mere change of variables in the functional integral \eqref{eq:Z}, it must leave it invariant. This then yields, using~\eqref{eq:defpsi}, the functional Ward identities
\begin{equation}\label{eq:xtshiftWI}
    \dfrac{\delta\Gamma{[\Psi]}}{\delta\Psi_i} = \dfrac{\delta S[\Psi]}{\delta\Psi_i}\,,\quad \Psi_i\in \Big\{ \la\vom\ra_{\cal J}, \la\tvom\ra_{\cal J}, \la\theta\ra_{\cal J}, \la\ttheta\ra_{\cal J}\big\}\,.
\end{equation}
Note that the corresponding global (non-infinitesimal) constant shifts are exact symmetries of the action, and the associated Ward identities have the same form as \eqref{eq:xtshiftWI}, but  integrated over space and time. Thus, the identities~\eqref{eq:xtshiftWI}, which are local both in space and time, are more constraining.
The identities~\eqref{eq:xtshiftWI} imply that the full dependence of $\Gamma$ on the mean values of the auxiliary fields $\vom, \tvom, \theta, \ttheta$ is known explicitly and keeps the bare form determined by the action. This means that this whole sector is not renormalized, and hence we use in the following the same notation for the fields $\varphi\in \{\vom, \tvom, \theta, \ttheta\}$ and their averages $\langle \varphi\rangle_{{\cal J}}$. In particular, these fields do not enter any $n$-point vertex function with $n>2$ since they appear only in quadratic terms in $\Gamma$.  As a consequence,  vertex functions with $n>2$ only involve velocity or response velocity fields, so that we set the notation
\begin{multline}\label{eq:vertexmn}
    \Gamma^{(m,n)}_{\alpha_1...\alpha_{m+n}}(t_1,\vx_1;\hdots;t_{m+n},\vx_{m+n}) \equiv
    \\
   \dfrac{ \delta^{m+n} \Gamma }
    {\delta u_{\alpha_1}(t_1,\vx_1)  \hdots
    \delta\tu_{\alpha_{m+n}}(t_{m+n},\vx_{m+n})}
    \Big|_{\vu, \tvu=0}\, ,
\end{multline}
where the first $m$ fields are average velocities $\vu =\langle \vv\rangle_{{\cal J}}$ and the $n$ last are average response velocities $\tvu =\langle \tvv\rangle_{{\cal J}}$.
Because of translational invariance in space and time, their Fourier transforms take the form
\begin{multline}\label{eq:vertexFourier}
  \Gamma^{(m,n)}_{\alpha_1...\alpha_{m+n}}(\omega_1,\vq_1;\cdots;\omega_{m+n},\vq_{m+n}) = \\(2\pi)^{d+1}\delta\left(\sum_{\ell=1}^{m+n}\omega_{\ell}\right)\delta^d\left(\sum_{\ell=1}^{m+n}\vq_{\ell}\right)\\
  \times\bar{\Gamma}^{(m,n)}_{\alpha_1...\alpha_{m+n}}(\omega_1,\vq_1;\cdots;\omega_{m+n-1},\vq_{m+n-1})  \, ,
\end{multline}
where the last frequency and momentum of $\bar{\Gamma}^{(m,n)}$ are implicit as they are fixed by the conservation of the total frequency and momentum.

\subsubsection{Time-gauged shift symmetry of the response field.}

The transformation
\begin{equation}
\label{eq:shift}
\begin{cases}
    \bar v_{\alpha}(t,\vx)
    \rightarrow
    \bar v_{\alpha}(t,\vx) + \bar \varepsilon_{\alpha}(t)
    \\
    \tom_{\alpha}(t,\vx)
    \rightarrow
    \tom_{\alpha}(t,\vx) + \epsilon_{\alpha \beta \gamma} \bar\varepsilon_{\beta}(t) v_{\gamma}(t,\vx) - \bar\varepsilon_{\alpha}(t) \theta(t,\vx)\,,
\end{cases}
\end{equation}
where $\bar{\mathbf{\varepsilon}}$ is a vectorial infinitesimal time-dependent shift is an extended symmetry of the Burgers action. By writing again that the transformation~\eqref{eq:shift} must leave  ${\cal Z}$ unaltered, one obtains the functional Ward identity
\begin{multline}\label{eq:tshiftWI}
    \int_{\vx} \left(
    \frac{\delta\Gamma}{\delta \tu_{\alpha}} -
    \epsilon_{\alpha\beta\gamma}\frac{\delta\Gamma}{\delta \tom_{\beta}}u_{\gamma} -
    \theta\frac{\delta\Gamma}{\delta \tom_{\alpha}}
    \right)
    =\\
    \int_{\vx} \left(
    \partial_t u_{\alpha} +
    \epsilon_{\alpha\beta\gamma}\partial_{\beta} w_{\gamma}
    \right)\,.
\end{multline}
Taking $m$ functional derivatives of this identity with respect to $\vu$ and $n$ with respect to $\tvu$, one obtains an identity for $\int_{\vx} \Gamma^{(m,n+1)}$ from the first term in the left-hand-side. Evaluating this identity at vanishing fields, one finds, using \ref{eq:xtshiftWI}, that this term vanishes. Because of the integration over space, this  yields in Fourier space that one of the momentum of this vertex associated with a response field is zero, such that one obtains:
\begin{align}\label{eq:tshiftWIvertex}
    \bar\Gamma^{(m,n+1)}_{\alpha_1\cdots\alpha_{m+n+1}} (\omega_1,\vp_1;\omega_2,\vp_2;\hdots,\underbrace{\,\omega_{k>m},0;\,}_{\bar{\mathbf{u}}}\hdots) = 0\, ,
\end{align}
for any $m,n$ except the lowest ones, which are given by the exact expressions:
\begin{align}
    \bar\Gamma^{(1,1)}_{\alpha_1\alpha} (\omega_1,0) &= i\omega_1\delta_{\alpha\alpha_1},
    \\
    \bar\Gamma^{(2,1)}_{\alpha_1\alpha_2\alpha} (\omega_1,\vp;\omega_2,-\vp) &= \delta_{\alpha\alpha_2} i p_{\alpha_1} - \delta_{\alpha\alpha_1} i p_{\alpha_2}\,,
\end{align}
where the identity~\eqref{eq:xtshiftWI} was used in the second expression.

\subsubsection{Time-gauged Galilean symmetry.}

The transformation
\begin{equation}
\label{eq:gal}
\begin{cases}
    v_{\alpha}(t,\vx) \rightarrow v_{\alpha}(t,\vx) - \partial_t \varepsilon_{\alpha}(t) + \varepsilon_{\beta}(t) \partial_{\beta} v_{\alpha}(t,\vx)\\
    \varphi(t,\vx) \rightarrow \varphi(t,\vx) + \varepsilon_{\beta}(t) \partial_{\beta} \varphi(t,\vx), \; \varphi \in \{\tvv, \vom, \tvom, \theta, \ttheta \}
\end{cases}
\end{equation}
where ${\mathbf{\varepsilon}}$ is a vectorial infinitesimal time-dependent function is an extended symmetry of the Burgers action. This reduces to the standard Galilean symmetry when $\varepsilon(t)$ is simply linear in $t$. Performing the change of variables \eqref{eq:gal} in \eqref{eq:Z}, one deduces the functional identity
\begin{equation}\label{eq:galileanWI}
    -\int_{\vx} \partial_t^2 \bar u_{\alpha} =  \int_{\vx}  \partial_t\frac{\delta\Gamma}{\delta u_{\alpha} }
  +  \sum_{
        \begin{smallmatrix}
        \varphi \in \{
        \vu, \tvu, \\
        \vom, \tvom,
        \theta, \ttheta
        \}
        \end{smallmatrix}
    }
   \int_{\vx} \partial_{\alpha} \varphi \frac{\delta\Gamma}{\delta \varphi}\,.
\end{equation}
Taking functional derivatives of this identity with respect to $m$ velocity and $n$  response velocity fields and evaluating the resulting identity at zero fields provides an exact relation for the  vertex function $\bar\Gamma^{(m+1,n)}$ with a zero momentum associated with a velocity field.  In Fourier space, it reads as
\begin{multline}\label{eq:galileanWIvertex}
    \bar\Gamma^{(m+1,n)}_{\alpha\alpha_1\hdots\alpha_{m+n}} (
    \underbrace{\,\omega,0;\,}_{\vu}\omega_1,\vp_1;\hdots;\omega_{m+n},\vp_{m+n}
    ) =
    \\
    -\sum_{j=1}^{m+n-1}
    \frac{p_{j\alpha}}{\omega}
    \left[
        \bar\Gamma^{(m,n)}_{\alpha_1\hdots\alpha_{m+n}}(
        \omega_1,\vp_1;
        \hdots;
        \omega_j+\omega,\vp_j;
        \hdots
        )-
    \right.
        \\
    \left.
        \bar\Gamma^{(m,n)}_{\alpha_1\hdots\alpha_{m+n}}(
        \omega_1,\vp_1;
        \hdots;
        \omega_j,\vp_j;
        \hdots
        )
    \right],
\end{multline}
for any $m,n \geq 1$ except the lowest one, given exactly by:
\begin{equation}
\label{eq:wardshift}
    \bar\Gamma^{(1,1)}_{\alpha\alpha_1} (\omega,0) = i \omega \delta_{\alpha\alpha_1}\,.
\end{equation}
As an example, these identities consist in two non-trivial relations for the 3-point vertex functions, which are
\begin{align}
    &\bar\Gamma^{(2,1)}_{\alpha\alpha_1\alpha_2} (\omega,0;\omega_1,\vp_1) =
   \nonumber \\
   & - \frac{p_{1\alpha}}{\omega}\left[
        \bar\Gamma^{(1,1)}_{\alpha_1\alpha_2}(\omega_1+\omega,\vp_1)
        -
        \bar\Gamma^{(1,1)}_{\alpha_1\alpha_2}(\omega_1,\vp_1)
    \right],
   \label{eq:gam21Gal} \\
    &\bar\Gamma^{(1,2)}_{\alpha\alpha_1\alpha_2} (\omega,0;\omega_1,\vp_1) =
  \nonumber  \\
   & - \frac{p_{1\alpha}}{\omega}\left[
        \bar\Gamma^{(0,2)}_{\alpha_1\alpha_2}(\omega_1+\omega,\vp_1)
        -
        \bar\Gamma^{(0,2)}_{\alpha_1\alpha_2}(\omega_1,\vp_1)
    \right]\,,
\end{align}
since any $\Gamma^{(m,0)}$ vanishes for causality reasons~\cite{Canet2011heq}, and Galilean invariance provides no constraint on $\Gamma^{(0,3)}$.

\section{Functional renormalization group formalism for the Burgers equation}
\label{sec:FRG}

\subsection{The Wetterich equation}
\label{sec:wetterich}

We briefly set the basis of the FRG formalism, referring the reader  to comprehensive introductions, such as~\cite{Berges2002,Kopietz2010,Delamotte2012,Dupuis2021}, for more details.
The path integral \eqref{eq:Z} encompasses the integration over all trajectories of the fluctuating fields $\Phi$, which are solutions of the stochastic equation (\ref{eq:stochEq}) for different realizations of the noise, and over all scales (or all modes) of these fields.
The idea of the functional renormalization group is based on the Wilsonian RG. Instead of integrating over all scales simultaneously, one integrates up to a certain RG scale $\kappa^{-1}$ (or, in momentum space, $\kappa$), thereby obtaining a modified theory. One then considers how this modified theory evolves under an infinitesimal change of the RG scale, and this gives rise to a differential equation, akin a dynamical system, where the role ot time is played by the RG scale, and the corresponding trajectories are in an abstract theory space, and connect a given microscopic model to its effective theory at large distance.
From the technical point of view, this is achieved in FRG through the introduction of a scale-dependent `weight' $\exp(- \Delta \mathcal{S}_{\kappa})$ in the functional integral to suppress ﬂuctuations below the RG scale $\kappa$:
\begin{equation}\label{Zkappa}
    \mathcal{Z}_{\kappa} = \int {\mathrm{D}} \Phi \exp \left(
        -\mathcal{S}[\Phi] - \Delta \mathcal{S}_{\kappa}[\Phi] + \int_{t,\vx} \mathcal{J} \Phi
    \right),
\end{equation}
where $\Delta \mathcal{S}_{\kappa}[\Phi] \equiv \frac{1}{2}\int_{t,\vx,\vx'} \Phi(t,\vx) \mathcal{R}_{\kappa}(|\vx-\vx'|) \Phi(t,\vx')$. The kernel $\mathcal{R}_{\kappa}$ is called `regulator' and must satisfy the following properties (in  Fourier space):
\begin{enumerate}
  \item $\mathcal{R}_{\kappa}(\vp) \stackrel{\kappa \rightarrow \infty}{\longrightarrow} \infty$: all fluctuations are frozen;
  \item $\mathcal{R}_{\kappa}(\vp) \stackrel{\kappa \rightarrow 0}{\longrightarrow} 0$: all fluctuations are averaged over;
   \item $\mathcal{R}_{\kappa}(\vp) \stackrel{|\vp| \ll \kappa}{\sim} \kappa^2$: the contribution of the `slow'

   \vspace{-0.cm}\hspace{2.5cm} modes is suppressed;
   \item $\mathcal{R}_{\kappa}(\vp) \stackrel{|\vp| \gg \kappa}{\longrightarrow} 0$: `fast' modes are unaltered

     \vspace{-0.cm}\hspace{2.5cm} and averaged over.
\end{enumerate}
The central object within the FRG formalism is the effective average action $\Gamma_\kappa$, defined as the following  modified Legendre transform of ${\cal W}_\kappa =\ln{\cal Z}_\kappa$
\begin{equation}
    \Gamma_{\kappa}[\Psi] + \Delta\mathcal{S}_{\kappa}[\Psi] =  \underset{\mathcal{J}}{\sup} \left[
        \int_{t,\vx}
        \mathcal{J} \Psi 
        -\mathcal{W}_{\kappa}[\mathcal{J}]
    \right]\,,
\end{equation}
where, as before, $\Psi\equiv\langle\Phi\rangle_{\cal J}$. In the large-$\kappa$ limit, the microscopic action $\Gamma_\infty\equiv \mathcal{S}$ is recovered since all the fluctuations are frozen, while in the small-$\kappa$ limit, the full effective action  $\Gamma_0\equiv\Gamma$, encompassing all the fluctuations at all scales, is obtained \cite{Delamotte2012,Dupuis2021}.  This provides $\Gamma_\kappa$ with a nice   physical meaning as the effective theory at scale $\kappa$, interpolating from the microscopic theory  at $\kappa=\Lambda$ to the complete effective theory when $\kappa\to 0$. The very idea of the Wilsonian RG is thus realized through this formalism in a very intuitive way.

For the Burgers equation, the macroscopic scale is set by the size of the system, or by the integral scale if the forcing acts at large scales. This scale can be denoted by $L$ (or $k_L=L^{-1}$ for momentum), and it embodies the scale beyond which ({\it i.e.} for $\kappa \lesssim k_L$) the RG flow essentially stops, because fluctuations are negligible or inexistent beyond this scale. The microscopic (momentum) scale, denoted $\Lambda$, defines the maximal momentum accessible in the system -- $\Lambda^{-1}$ can be thought of as the scale at which the continuous fluid description is valid.

The evolution of $\Gamma_{\kappa}$ with the RG scale in between these two
limits is given by the exact Wetterich equation~\cite{Wetterich93,Ellwanger94,Morris94}
\begin{equation}\label{Wetterich}
    \partial_{\kappa} \Gamma_{\kappa} =
    \frac{1}{2}\textrm{tr}\,\int_{\omega,\vq}
    \partial_{\kappa} \mathcal{R}_{\kappa}\,
    \mathcal{G}_{\kappa},
\end{equation}
where the  trace means summation over all fields, and
\begin{align}\label{G}
\mathcal{G}_{\kappa} \equiv \left(
    \Gamma_{\kappa}^{(2)} + \mathcal{R}_{\kappa}
\right)^{-1}
\end{align}
is the propagator matrix, where the Hessian
\begin{align}\label{Gamma2}
    \Gamma_{\kappa}^{(2)}(t_1,\vx_1;t_2,\vx_2)  \equiv \left. \frac
    {\delta^{2} \Gamma}
    {
    \delta\Psi(t_1,\vx_1)
    \delta\Psi(t_2,\vx_2)
    }
    \right|_{\Psi=0}
\end{align}
is the  matrix of two-point vertex functions. Their exact flow can be obtained by taking functional derivatives of Eq.~(\ref{Wetterich}), yielding in Fourier space:
\begin{align}
\p_s & \bar{\Gamma}_{\kappa,AB}^{(2)}(\varpi,\vp)  = \dfrac 1 2 {\rm tr} \Bigg[\tilde{\p}_s  \int_{\omega,\vq} \bar{\cal G}_{\kappa}(\omega,\vq) \label{eq:dsgam2}\\
&\cdot \Big\{ \bar{\Gamma}_{\kappa,AB}^{(4)}(\varpi,\vp,-\varpi,-\vp,\omega,\vq) -  \bar{\Gamma}_{\kappa,A}^{(3)}(\varpi,\vp,\omega,\vq)\nonumber\\
 &\cdot \bar{\cal G}_\kappa(\varpi+\omega,\vp+\vq)\cdot \bar{\Gamma}_{\kappa,B}^{(3)}(-\varpi,-\vp,\varpi+\omega,\vp+\vq)\Big\}\, \Bigg],\nonumber
\end{align}
where  the operator $\tilde{\p}_s$ only acts on the regulators, {\it i.e.} $\tilde{\p}_s\equiv \p_s {\cal R}_{\kappa,AB}\frac{\p}{\p {\cal R}_{\kappa,AB}}$, with $s=\ln(\kappa/\Lambda)$  the RG `time', and $\p_s=\kappa\p_\kappa$. The 3- and 4-point vertices $\bar{\Gamma}_{\kappa,A}^{(3)}$ and $\bar{\Gamma}_{\kappa,AB}^{(4)}$ are written as $6\times 6$ matrices, the remaining fields being fixed to the external ones (labeled by the indices $A$ and $B$ from the multiplet $\Psi$). The summation is carried over the implicit field indices (matrix product), and integration  over the internal frequency and momentum  $(\omega,\vq)$.

\subsection{Structure of the propagator matrix}

One can infer from the set of Ward identities (\ref{eq:xtshiftWI}) the general form of the effective average action as
\begin{align}
&\Gamma_{\kappa}[\vu,\tvu,\vom,\tvom,\theta,\ttheta] = \tilde\Gamma_{\kappa}[\vu, \tvu] + \label{eq:Gammakappa}\\
&\int_{t,\vx} \Big\{
   \tvu \cdot (\nabla \times \vom)  + \tvom \cdot (\nabla \times \vu)
    + \ttheta \nabla \cdot \vom   + \tvom \cdot \nabla \theta
\Big\}\,,\nonumber
\end{align}
where the explicit part of $\Gamma_{\kappa}$ containing the auxiliary fields is not renormalized, and $\tilde\Gamma_{\kappa}$ does not depend on them.
One can then deduce by taking two functional derivatives of \eqref{eq:Gammakappa}  the general form of the Hessian of $\Gamma_\kappa$, and, by inverting this Hessian, the general form of the propagator matrix ${\cal G}_\kappa$, which are both given in Appendix~\ref{app:prop}.
Since the auxiliary-field sector does not renormalize, the RG procedure of progressive integration of fluctuation modes only needs to be carried out on the velocity and response velocity fields.
Thus, one can choose a regulator matrix $\mathcal{R}_{\kappa}$ with only three non-zero elements ${\cal R}_{\kappa,\vu \tvu}={\cal R}_{\kappa,\tvu \vu}\equiv {\cal M}_\kappa$ and ${\cal R}_{\kappa,\tvu \tvu}\equiv {\cal N}_\kappa$ (${\cal R}_{\kappa,\vu \vu}$ is fixed to ${\cal R}_{\kappa,\vu \vu}=0$ to preserve causality along the flow). Moreover, since the auxiliary fields do not enter any $n>2$ vertex function, they completely decouple from the flow equations~\eqref{eq:dsgam2}. One can thus focus on the sole velocity and response velocity sector and consider  the reduced vertex functions~\eqref{eq:vertexmn}, and the reduced propagator matrix $\bar{G}_\kappa$  given by (see Appendix~\ref{app:prop})
\begin{equation}
    \bar{G}_{\kappa}(\omega,\vp) \equiv
    \begin{blockarray}{ccc}
      \quad\quad\quad\quad        & u_\beta & \tu_\beta \\
\begin{block}{c(cc)}
  u_\alpha   &\bar{G}^{(2,0)}_{\kappa,\alpha\beta}(\omega,\vp) & \bar{G}^{(1,1)}_{\kappa,\alpha\beta}(\omega,\vp) \\
  \tu_\alpha  &\bar{G}^{(1,1)}_{\kappa,\alpha\beta}(-\omega,\vp) & 0\\
    \end{block}
     \end{blockarray}
\end{equation}
where
\begin{align}
   \bar{G}^{(1,1)}_{\kappa,\alpha\beta}(\omega,\vp)  &\equiv
    \dfrac
    {1}
    {
     \bar\Gamma^{(1,1)}_{\kappa,\parallel}(\omega,\vp) + {\cal M}_{\kappa}(\vp)
    } 
     P^{\parallel}_{\alpha\beta}(\vp),\label{eq:G11} \\
    \bar{G}^{(2,0)}_{\kappa,\alpha\beta}(\omega,\vp)  &\equiv
    \dfrac
    {-(\bar\Gamma^{(0,2)}_{\kappa,\parallel}(\omega,\vp) +{\cal N}_{\kappa}(\vp))}
    {\left|
    \bar\Gamma^{(1,1)}_{\kappa,\parallel}(\omega,\vp) + {\cal M}_{\kappa}(\vp)
    \right|^2} 
     P^{\parallel}_{\alpha\beta}(\vp)\,,\label{eq:G20}
\end{align}
with the longitudinal projector
\begin{equation}
\label{eq:proj}
    P^{\parallel}_{\alpha\beta}(\vp) = \frac{p_{\alpha} p_{\beta}}{p^2}\,,
\end{equation}
and where $\bar{\Gamma}_{\kappa,\parallel}^{(1,1)}$ is the longitudinal component of $\Gamma^{(1,1)}_{\kappa,\alpha\beta}$ defined as $\bar{\Gamma}_{\kappa,\parallel}^{(1,1)}(\omega,\vq)=P_{\alpha\beta}^\parallel(\vq) \bar\Gamma^{(1,1)}_{\kappa,\alpha\beta}(\omega,\vq)$, and similarly for $\bar{\Gamma}_{\kappa,\parallel}^{(0,2)}$.
 The reduced  propagator  matrix is purely longitudinal for the Burgers equation as a consequence of irrotationality, whereas it is purely transverse  for the Navier-Stokes equation as a consequence of incompressibility \cite{Canet2016, Tarpin2018}.

The exact flow equations~\eqref{eq:dsgam2} can be projected onto the reduced $(\vu,\tvu)$ sector.
 The different terms involved in these flow equations  can be written as diagrams, representing the field $\vu$ by a line carrying an ingoing-into-a-vertex arrow and the response field $\tvu$ by a line carrying an outgoing-of-a-vertex arrow.
Performing the matrix products and the trace  for $A=B=\tvu$ leads to the exact flow equation for $\Gamma_\kappa^{(0,2)}$ represented diagrammatically on Fig.~\ref{fig:flowG02} (without the combinatorial factors).
\begin{figure}[h]
\begin{tikzpicture}
 \path [draw=black,postaction={on each segment={mid arrow=black}}]
  (-0.5,0.1) -- (0,-0.35)
  (-0.5,0.1) -- (-1,-0.35)
  ;
  \draw (-0.5,.5) circle(.4);
   \node at (-0.5,-0.6) {  $(a)$};
    \node at (-0.5,0.9) {\scriptsize $\bullet$};
  \path [draw=black,postaction={on each segment={mid arrow=black}}]
  (-0.9,0.45) -- (-0.9,0.44)
  (-0.1,0.45) -- (-0.1,0.44);
   \node at (-2.5,0.2) {   $\p_s\Gamma_\kappa^{(0,2)} = \quad \tilde\p_s\Bigg\{\;$};
    \node at (0.5,0.2) {  $+$};
  \path [draw=black,postaction={on each segment={mid arrow=black}}]
    (1.5,0.1) -- (2.,-0.35)
    (1.5,0.1) -- (1.,-0.35)
    ;
  \draw (1.5,0.5) circle(.4);
   \node at (1.5,-0.6) {  $(b)$};
  \path [draw=black,postaction={on each segment={mid arrow=black}}]
  (1.1,0.45) -- (1.1,0.44)
  (1.9,0.54) -- (1.9,0.55);
   \node at (2.5,0.2) {  $-$};
    \path [draw=black,postaction={on each segment={mid arrow=black}}]
  (4.4,0.2) -- (5.,0.2)
  (3.6,0.2) -- (3.,0.2)
  ;
  \draw[color=black] (4.,0.2) circle(.4);
   \node at (4.,-0.6) {  $(c)$};
   \node at (4.,0.6) {\scriptsize $\bullet$};
    \node at (4.,-0.2) {\scriptsize $\bullet$};
   \path [draw=black,postaction={on each segment={mid arrow=black}}]
  (4.35,0.41) -- (4.352,0.408)
  (3.65,0.41) -- (3.648,0.408)
  (3.65,-0.01) -- (3.648,-0.008)
  (4.35,-0.01) -- (4.352,-0.008)
  ;
   \node at (-3.5,-1.4) {  $-$};
   \path [draw=black,postaction={on each segment={mid arrow=black}}]
  (-2.4,-1.4) -- (-3,-1.4)
  (-1.6,-1.4) -- (-1,-1.4);
   \draw (-2,-1.4) circle(.4);
     \node at (-2,-2.2) {  $(d)$};
    \path [draw=black,postaction={on each segment={mid arrow=black}}]
    (-1.73,-1.11) -- (-1.732,-1.108)
    (-1.665,-1.63) -- (-1.663,-1.628)
    (-2.25,-1.71) -- (-2.248,-1.712)
   (-2.33,-1.17) -- (-2.332,-1.172)
  ;
    \node at (-0.5,-1.4) {  $-$};
      \path [draw=black,postaction={on each segment={mid arrow=black}}]
     (0.6,-1.4) -- (0,-1.4)
  (1.4,-1.4) -- (2,-1.4);
   \draw (1,-1.4) circle(.4);
    \node at (1,-2.2) {  $(e)$};
     \path [draw=black,postaction={on each segment={mid arrow=black}}]
      (1.27,-1.11) -- (1.268,-1.108)
    (1.27,-1.69) -- (1.268,-1.692)
    (0.67,-1.63) -- (0.668,-1.628)
   (0.67,-1.17) -- (0.668,-1.172)
  ;
    \node at (2.5,-1.4) {  $-$};
       \path [draw=black,postaction={on each segment={mid arrow=black}}]
     (3.5,-1.4) -- (2.9,-1.4)
  (4.3,-1.4) -- (4.9,-1.4);
   \draw (3.9,-1.4) circle(.4);
    \node at (3.9,-2.2) {  $(f)$};
      \path [draw=black,postaction={on each segment={mid arrow=black}}]
      (4.23,-1.17) -- (4.232,-1.172)
    (4.17,-1.69) -- (4.168,-1.692)
    (3.57,-1.63) -- (3.568,-1.628)
   (3.57,-1.17) -- (3.568,-1.172)
  ;
   \node at (3.9,-1) {\scriptsize $\bullet$};
    \node at (5.2,-1.4) {  $\Bigg \}$};
  \end{tikzpicture}
  \caption{Diagammatic representation of the flow equation of $\Gamma_\kappa^{(0,2)}$, which is given by the component $A=B=\tvu$ of the exact flow equation \eqref{eq:dsgam2}. The different diagrams are labeled by latin letters for use in the explicit calculations detailed in App.~\ref{app:IBfixed-point} and~\ref{app:largep}.}
  \label{fig:flowG02}
  \end{figure}

  Similarly, performing the matrix products and the trace in \eqref{eq:dsgam2} for $A=\vu$ and $B=\tvu$ leads to the exact flow equation for $\Gamma_\kappa^{(1,1)}$ represented diagrammatically on Fig.~\ref{fig:flowG11}.
  \begin{figure}[h]
\begin{tikzpicture}
 \path [draw=black,postaction={on each segment={mid arrow=black}}]
  (0,0.1) -- (0.5,-0.35)
  (-0.5,-0.35) -- (0,0.1)
  ;
  \draw (0,.5) circle(.4);
   \node at (0,-0.6) {  $(a)$};
    \node at (0,0.9) {\scriptsize $\bullet$};
  \path [draw=black,postaction={on each segment={mid arrow=black}}]
  (-0.4,0.45) -- (-0.4,0.44)
  (0.4,0.45) -- (0.4,0.44);
   \node at (-2,0.2) {   $\p_s\Gamma_\kappa^{(1,1)} = \quad \tilde\p_s\Bigg\{ \; $};
    \node at (1,0.2) {  $+$};
  \path [draw=black,postaction={on each segment={mid arrow=black}}]
    (2,0.1) -- (2.5,-0.35)
    (1.5,-0.35) -- (2,0.1)
    ;
  \draw (2,0.5) circle(.4);
   \node at (2,-0.6) {  $(b)$};
  \path [draw=black,postaction={on each segment={mid arrow=black}}]
  (1.6,0.45) -- (1.6,0.44)
  (2.4,0.54) -- (2.4,0.55);
   \node at (3,0.2) {  $-$};
    \path [draw=black,postaction={on each segment={mid arrow=black}}]
  (4.9,0.2) -- (5.5,0.2)
  (3.5,0.2) -- (4.1,0.2)
  ;
  \draw[color=black] (4.5,0.2) circle(.4);
   \node at (4.5,-0.6) {  $(c)$};
   \node at (4.5,0.6) {\scriptsize $\bullet$};
   \path [draw=black,postaction={on each segment={mid arrow=black}}]
  (4.85,0.41) -- (4.852,0.408)
  (4.15,0.41) -- (4.148,0.408)
  (4.228,-0.088) -- (4.23,-0.09)
  (4.85,-0.01) -- (4.852,-0.008)
  ;
   \node at (-2.,-1.4) {  $-$};
   \path [draw=black,postaction={on each segment={mid arrow=black}}]
   (-1.5,-1.4) -- (-0.9,-1.4)
  (-0.1,-1.4) -- (0.5,-1.4);
   \draw (-0.5,-1.4) circle(.4);
     \node at (-0.5,-2.2) {  $(d)$};
    \path [draw=black,postaction={on each segment={mid arrow=black}}]
    (-0.23,-1.11) -- (-0.232,-1.108)
    (-0.165,-1.63) -- (-0.163,-1.628)
    (-0.75,-1.71) -- (-0.748,-1.712)
   (-0.83,-1.17) -- (-0.832,-1.172)
  ;
 \node at (1.,-1.4) {$-$};
   \path [draw=black,postaction={on each segment={mid arrow=black}}]
   (1.5,-1.4) -- (2.1,-1.4)
  (2.9,-1.4) -- (3.5,-1.4);
   \draw (2.5,-1.4) circle(.4);
     \node at (2.5,-2.2) {  $(e)$};
    \path [draw=black,postaction={on each segment={mid arrow=black}}]
    (2.85,-1.19) -- (2.852,-1.192)
    (2.835,-1.63) -- (2.837,-1.628)
    (2.25,-1.71) -- (2.252,-1.712)
   (2.25,-1.09) -- (2.252,-1.088)
  ;
    \node at (4.,-1.4) {  $\Bigg \}$};
  \end{tikzpicture}
  \caption{Diagammatic representation of the flow equation of $\Gamma_\kappa^{(1,1)}$, which is given by the component $A=\vu$, $B=\tvu$ of the exact flow equation \eqref{eq:dsgam2}. The different diagrams are labeled by latin letters for use in the explicit calculations detailed in App.~\ref{app:IBfixed-point} and~\ref{app:largep}.}
  \label{fig:flowG11}
  \end{figure}

These flow equations are exact, but they are not closed in the sense that they involve higher-order 3- and 4-point vertices. 
We first solve them in Sec.~\ref{sec:LPA} within a simple approximation, in order to determine the  fixed-point structure of the Burgers-KPZ equation.
 We then stress in  Sec.~\ref{sec:largep} that in the inviscid limit, the regime of interest is the large momentum regime, and we show that the FRG flow equations can be closed exactly in the limit of large momentum thanks to the set of Ward identities~\eqref{eq:tshiftWIvertex} and~\eqref{eq:galileanWIvertex}.

\section{Fixed point structure of the Burgers-KPZ equation}
\label{sec:LPA} 

Our aim here is to provide the complete phase diagram of the Burgers-KPZ equation in all $d$, elucidating in particular the fate of the IB fixed point, which was only shown to exist in $d=1$ so far.
Let us recall the known behavior of this equation, in the language of the KPZ interface. For $d\leq 2$, the interface always roughens, and becomes scale-invariant at large distances, exhibiting a power-law behavior characterized by anomalous exponents. In terms of RG, this means that the rough phase is controlled by a non-Gaussian, fully attractive, fixed point, the celebrated KPZ fixed point. In $d>2$, a phase transition, called roughening transition (RT), occurs, depending on the microscopic  non-linearity $\lambda$. For $\lambda<\lambda_c$, the interface remains smooth, it is described by the Gaussian EW fixed point, whereas for $\lambda>\lambda_c$, it becomes rough. The phase transition is continuous, controlled by a non-Gaussian fixed point with one relevant (unstable) direction, akin the Wilson-Fisher fixed point in equilibrium Ising or ${\cal O}(N)$ models. The difference compared to  equilibrium is that not only the transition, but also the rough phase itself is controlled by a critical non-trivial fixed point, the KPZ fixed point, which is not the case for the fixed points associated with the ordered or the disordered phases in ferromagnets.  

Let us now investigate the complete phase diagram using the FRG formalism. The FRG flow equations are exact but they are functional partial differential equations which cannot be solved exactly in most cases (see, however, \cite{Benitez13}). Several approximation schemes to exploit them have been developed and thoroughly studied. We refer the readers to Refs.~\cite{Berges2002,Delamotte2012,Dupuis2021} for a detailed account of the standard approximations.
In this section, we use a very simple approximation, which consists in keeping the bare form of the action, but with effective flowing parameters, which means that the original parameters ${\cal D}$ and $\nu$ of the Burgers-KPZ action~\eqref{eq:action} are promoted to $\kappa$-dependent effective parameters ${\cal D}_\kappa$ and $\nu_\kappa$.
This amounts to defining the following ansatz for $\tilde\Gamma_\kappa$ defined in Eq.~\eqref{eq:Gammakappa}:
\begin{align}
\tilde\Gamma_\kappa[\vu,\tvu] =& \int_{t,\vx} \Big\{  \tvu \cdot \left[
          \mu_\kappa  \partial_t \vu + \lambda_\kappa (\vu \cdot \nabla) \vu - \nu_\kappa \nabla^2 \vu
        \right] \nonumber\\
        &- {\cal D}_\kappa (\nabla \cdot \tvu)^2 \Big\}\, ,
\label{eq:anzLPA}
\end{align}
where we also introduced the new parameters $\mu_\kappa$ and $\lambda_\kappa$ since the corresponding terms could in principle be renormalized.

Despite its simplicity, it was shown that this ansatz enables one to capture  the KPZ strong-coupling fixed point in all dimensions $d$ \cite{Canet2005b}, and also the IB fixed point in $d=1$ \cite{Fontaine2023InvBurgers}. It is thus sufficient to provide a {\it qualitative} picture of the phase diagram.  Of course, in order to obtain {\it quantitative} estimates of the critical exponents and scaling functions, one needs to resort to improved levels of approximation, such as the ones introduced in     Refs.~\cite{Canet2010,Canet2011kpz,Kloss2012}.

The advantage of the approximation \eqref{eq:anzLPA} is that it leads to  very simple flow equations. They are obtained by projecting the exact flow equations \eqref{eq:dsgam2} onto the reduced functional space defined by the ansatz \eqref{eq:anzLPA}.
In particular,   the associated two-point vertex functions are simply given by
\begin{equation}
 \bar{\Gamma}_{\kappa,\parallel}^{(1,1)}(\omega,\vq) = i \mu_\kappa \omega +\nu_\kappa \vq^{2} \,,\qquad \bar{\Gamma}_{\kappa,\parallel}^{(0,2)}(\omega,\vq) = -2 \vq^{2} {\cal D}_\kappa\, ,
 \label{eq:gam2LPA}
\end{equation}
 The only non-vanishing $n$-point vertex function with $n>2$ is the one present in the original action which reads as
\begin{equation}
 \bar{\Gamma}_{\kappa,\alpha\beta\gamma}^{(2,1)}(\omega_1,\vq_1,\omega_2,\vq_2) = -i \lambda_\kappa ( q_1^\beta \delta_{\alpha\gamma} + q_2^\alpha \delta_{\beta\gamma} ) \,. \label{eq:gam21}
\end{equation}
One can show that the shift symmetry (through the Ward identity \eqref{eq:wardshift}) imposes that $\mu_\kappa$ is not renormalized, {\it i.e.}  $\mu_\kappa \equiv  1$ at all scales. Similarly, the Galilean invariance (through the Ward identity \eqref{eq:gam21Gal}) imposes that $\lambda_\kappa$ is not renormalized \cite{Frey94,Canet2011kpz}. We keep the notation $\lambda$ for this constant ($\lambda\equiv1$) for convenience, and $\lambda_\kappa \equiv  \lambda$ at all scales \cite{Frey94,Canet2011kpz}.
The flows of $\nu_\kappa$ and ${\cal D}_\kappa$ can be obtained as
\begin{align}
 \partial_s {\cal D}_\kappa &= -\dfrac {1}{4d}\p_{p_\alpha}\p_{p_\alpha}  \partial_s  \bar{\Gamma}_{\kappa,\parallel}^{(0,2)}(\varpi,\vp)\Big|_{\varpi=0,\vp=0}\,\nonumber\\
 \partial_s \nu_\kappa &= \dfrac {1}{2d} \p_{p_\alpha}\p_{p_\alpha}\Re{\rm e} \left[\partial_s  \bar{\Gamma}_{\kappa,\parallel}^{(1,1)}(\varpi,\vp)\Big|_{\varpi=0,\vp=0}\right]\,.
 \label{eq:defNu}
\end{align} 
The calculation is detailed in  Appendix~\ref{app:IBfixed-point}.

\begin{figure}
\includegraphics[width=7cm]{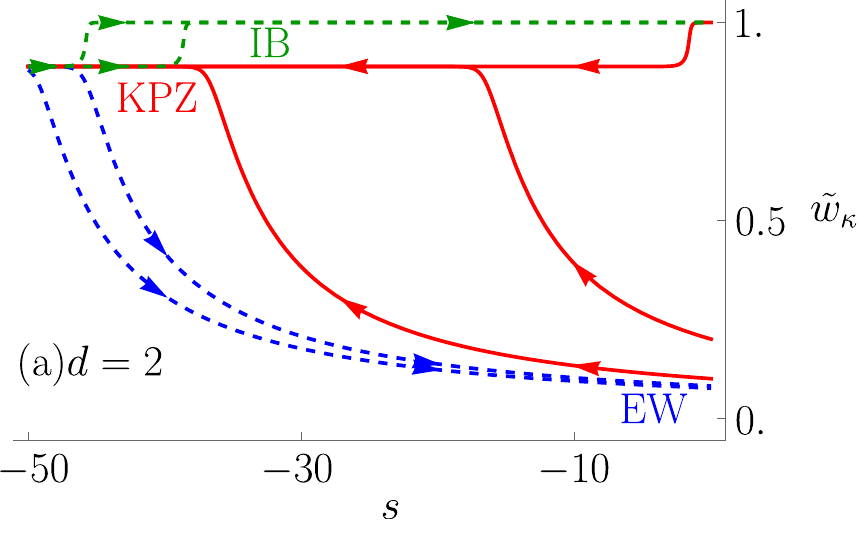}
\vspace{0.5cm}
\includegraphics[width=7cm]{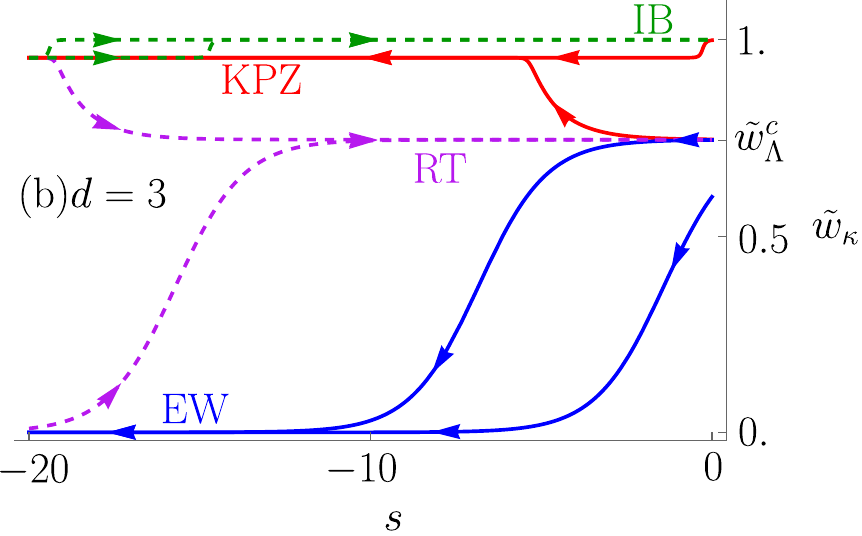}
 \caption{FRG flows of $\tilde{w}_\kappa$ in $d = 2$ (a) and in $d=3$ (b). Plain lines are IR flows from $s=0$ ($\kappa=\Lambda$) towards $s\to-\infty$ ($\kappa\to 0$). Several typical trajectories from different initial conditions are represented, the ones with the same color lead to the same large-distance behavior, which is a hallmark of universality.
 For $d\leq 2$ (a), the IR flow (red arrows) always leads to the KPZ fixed point, for any initial condition $0<\tilde{w}_\Lambda<1$. For $d> 2$ (b), the IR flow either leads  to the KPZ fixed point (red arrows) for all  $\tilde{w}^c_\Lambda<\tilde{w}_\Lambda<1$ or to the EW fixed point (blue arrows) for  all $0<\tilde{w}_\Lambda<\tilde{w}^c_\Lambda$.
  Dashed lines are UV flows from a large negative value of  $s_*(d)$  (small $\kappa_*(d)$) -- defined in the text below \eqref{eq:gtilde} -- towards $s=0$ ($\kappa=\Lambda$). Several typical trajectories are again represented. For $d\leq 2$  (a), the flow either leads to   the EW fixed point (blue arrows) for all  $0<\tilde{w}_{\kappa_*(d)}<\tilde{w}^{\rm KPZ}_*$ or to the IB fixed point (green arrows) for all $\tilde{w}^{\rm KPZ}_*<\tilde{w}_{\kappa_*(d)}<1$. For $d>2$ (b), the flow still leads to the IB fixed point (green arrows) for all $\tilde{w}^{\rm KPZ}_*<\tilde{w}_{\kappa_*(d)}<1$, but it now leads to the RT fixed point (purple arrows) for all $0<\tilde{w}_{\kappa_*(d)}<\tilde{w}^{\rm KPZ}_*$. Thus there are two UV stable (IR unstable) fixed points in all $d$, IB and EW or RT. }
 \label{fig:flowFRG}
\end{figure}

In order to study the fixed points, one introduces dimensionless quantities, denoted with a hat symbol, such that the effective average action \eqref{eq:anzLPA} expressed in terms of these quantities becomes $\kappa$-independent. One can check that, measuring momenta in units of $\kappa$, the appropriate  rescaling is
\begin{equation}
\label{eq:rescaling}
 \hat{\vq} = \dfrac{\vq}{\kappa}\, , \; \hat{\omega} = \dfrac{\omega}{\kappa^2\nu_\kappa}\, ,\; \hat{\tvu} = \sqrt{\frac{\cal D_\kappa}{\kappa^{d}\nu_\kappa}}\tvu\, ,\; \hat{\vu} = \sqrt{\frac{\nu_\kappa}{\kappa^{d}{\cal D}_\kappa}}\vu.
\end{equation}
In a scaling regime, the effective noise amplitude ${\cal D}_\kappa$ and viscosity $\nu_\kappa$ are expected to behave as power-laws ${\cal D}_\kappa\sim \kappa^{-\eta^{\cal D}}$ and  $\nu_\kappa\sim \kappa^{-\eta^\nu}$ with some anomalous exponents $\eta^{\cal D}$ and $\eta^{\nu}$. This leads one to  define the two effective anomalous dimensions
\begin{equation}
\eta_\kappa^\nu = -\p_s \ln \nu_\kappa\,,\qquad\eta_\kappa^{\cal D} = -\p_s \ln {\cal D}_\kappa\,.
\end{equation} 
If the flow reaches a fixed point, the effective anomalous exponents acquire fixed values, denoted with an asterisk  $\eta_\kappa^{\cal D}\to \eta_*^{\cal D}$ and $\eta_\kappa^\nu\to \eta_*^\nu$ when $\kappa\to 0$ (which then identify with the anomalous exponents $\eta^{\cal D}$ and $\eta^{\nu}$ previously introduced).

The  physical roughness and dynamical exponents  $\chi$ and $z$ can be defined from the  (dimensionful) velocity correlation function, which is expected to endow in a scaling regime the following scaling form~\cite{Kardar86,Halpin-Healy95}
\begin{equation}
 \bar{C}(\omega,\vq) \equiv \bar{G}_{\parallel}^{(2,0)}(\omega,\vq) = |\vq|^{2-d-2\chi-z} F(\omega/|\vq|^z)\, ,
\end{equation}
where  $F$ is a universal scaling function.
From the rescaling \eqref{eq:rescaling}, one deduces that the exponents  $\chi$ and $z$ are related  to the anomalous exponents as \cite{Kloss2012}
\begin{equation}
\label{eq:defexpo}
z=2-\eta_*^\nu \,,\qquad \chi = \dfrac 1 2(2-d +\eta_*^{\cal D}-\eta_*^\nu)\,.
\end{equation}
The dimensionless effective coupling, associated with the non-linear term, is then defined as $\hat g_\kappa = \kappa^{d-2}\lambda^2 {\cal D}_\kappa/\nu_\kappa^3 $. Its flow equation is thus given by
\begin{equation}
\p_s \hat g_\kappa = \hat g_\kappa\big(d-2 -\eta_\kappa^{\cal D}+3 \eta_\kappa^\nu \big)\, .
\end{equation} 
In fact, in order to study the IB fixed point, which corresponds to $\hat g_*\to \infty$, it is convenient to change variables to
\begin{equation}
 \hat w_\kappa = \hat g_\kappa/(1+\hat g_\kappa)\, ,
\end{equation}
whose flow equation reads as
\begin{equation}
\label{eq:dsw}
\p_s \hat w_\kappa = \hat w_\kappa(1-\hat w_\kappa)\big(d-2 -\eta_\kappa^{\cal D}+3 \eta_\kappa^\nu \big)\, .
\end{equation} 
Let us discuss the solutions of the fixed-point equation $\p_s \hat w_\kappa =0$.
There are three ways to satisfy this equation: $\hat w_*=0$, $\hat w_*=1$, or $(d-2 -\eta_*^{\cal D}+3 \eta_*^\nu )=0$. This last equation yields, using the definition \eqref{eq:defexpo} of the critical exponents, the identity $z+\chi=2$ valid in any $d$. There are two such solutions with  $0<\hat{w_*}<1$ and  $z+\chi=2$, which are the KPZ fixed point and the roughening transition (RT) fixed point.  In contrast,  this relation is not imposed neither for $\hat w_*^{\rm EW}=0$, which is the EW fixed point, nor for $\hat w_*^{\rm IB}=1$, which is the IB fixed point (although Galilean symmetry is preserved at both these fixed points).
In these cases, the actual fixed-point values of the anomalous dimensions $\eta_*^\nu$ and $\eta_*^{\cal D}$ cannot be read off from the fixed point equation for $\hat w_\kappa$ alone, but one has to solve the full set of equations \eqref{eq:dsw}, \eqref{eq:etanuD}.

We numerically solved these equations for dimensions $0\leq d\leq 6$.
Note that in the numerics we absorbed the factor $v_d=(2^{d-1}\pi^{d/2}\Gamma(d/2))^{-1}$ coming from the $d$-dimensional angular integration into the coupling, defining
\begin{equation}
\label{eq:gtilde}
   \tilde{g}_\kappa\equiv  \hat{g}_\kappa v_d\, ,\quad \tilde w_\kappa= \tilde{g}_\kappa/(1+\tilde{g}_\kappa)\,.
\end{equation}
 The plots are displayed using either $\tilde{w}$ or $\tilde{g}$ for clarity purpose.
In each dimension, we first integrated the flow from $s=0$ ($\kappa=\Lambda$) with different initial conditions $\hat w_\Lambda$ towards the IR $s\to-\infty$ ($\kappa\to0$) until a fixed point was reached, stopping at a value that we denote $s_*(d)$ (or equivalently $\kappa_*(d)$). As examples, in Fig.~\ref{fig:flowFRG} these values are $s_*(2)\simeq -20$ and  $s_*(3)\simeq -50$. For all initial conditions $0<\hat w_\Lambda<1$, the  fixed point reached  is always the KPZ one for $d\leq 2$.  This shows that this fixed point is fully attractive, corresponding to an interface which is always rough as expected physically. In  $d>2$, the fixed point reached as $s\to-\infty$ is either the KPZ fixed point, when $\hat w_\Lambda> \hat w_\Lambda^c$, or the EW fixed point, when $\hat w_\Lambda< \hat w_\Lambda^c$, where $\hat w_\Lambda^c$ is the critical value for the roughening transition. The RT fixed point has one relevant direction, which means that the value of $\hat w_\Lambda$ has to be fine-tuned to $\hat w_\Lambda^c$, which is akin a critical temperature, to be at the transition. We refer to this  flow from small scales to large scales as the  IR flow. This corresponds to the standard Wilsonian flow.  The fixed points reached in the IR flow (KPZ in $d<2$, KPZ or EW in $d\geq 2$) are termed IR stable, since they are attractive as the flow runs towards the large distances. The EW fixed point for $d\leq 2$ and the RT fixed point are IR unstable.

In order to study the IR unstable fixed points, one can run the flow
   backwards, starting at a scale $s\simeq {s_*(d)}$ from initial conditions
  $\hat w_{\kappa_*(d)}\lesssim \hat w_*^{\rm KPZ}$ and $\hat w_{\kappa_*(d)}\gtrsim \hat w_*^{\rm KPZ}$ and letting $s\to0$. We refer to this flow from large scales to small scales as the UV flow. Since the sign of the evolution in the RG time is opposite compared to the  IR flow, the stability of the fixed points is reversed. Hence, the IR stable fixed points become unstable in this flow, and IR unstable fixed point become attractive in this reversed flow, {\it i.e.} UV stable. Of course, the UV flow does not correspond to the standard RG program to build a large-scale effective theory, we use it simply as a convenient way to reach the IR unstable fixed points.  Numerically integrating  the reversed flow, we found, as UV stable fixed points,  the IB and EW fixed points for $d\leq 2$, and the IB and RT fixed points for $d>2$. This shows the existence of the IB fixed point in all $d$, at least within the crude approximation~\eqref{eq:anzLPA}.
  We show in Fig.~\ref{fig:flowFRG} the IR and UV flows in $d=2$ as an illustration of the $d\leq 2$ case, and the IR and  UV flows in $d=3$ as an illustration of the $d>2$ case.

\begin{figure}
\includegraphics[width=8cm]{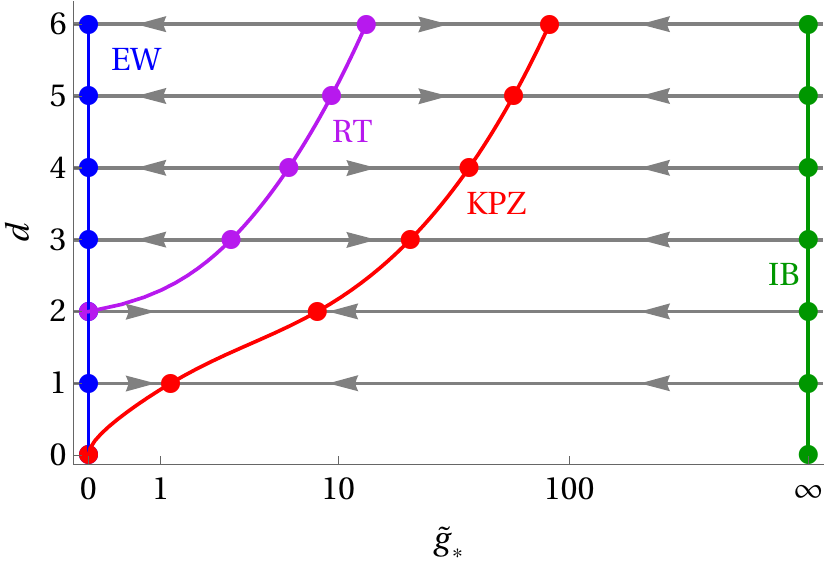}
 \caption{Flow diagram of the Burgers-KPZ equation:  fixed point values of $\tilde g_*$ as a function of the dimension $d$. The dots represent the  fixed points, and the arrows indicate IR flows (from small scales to large scales).  The KPZ fixed point is thus always attractive, and the RT and IB fixed points always unstable. The EW fixed point changes stability in $d=2$, from unstable in $d\leq 2$ to stable in $d>2$.}
 \label{fig:flowdiag}
\end{figure}

  To summarize, within the approximation~\eqref{eq:anzLPA}, there are four fixed points:
\begin{itemize}
 \item EW fixed point: $\hat w_*^{\rm EW}=0$,

\nd IR stable for $d>2$, and UV stable for $d\leq 2$.

\item IB fixed point:  $\hat w_*^{\rm IB}=1$, always UV stable.

\item  KPZ fixed point: $0<\hat w_*^{\rm KPZ}<1$, always IR stable.

\item RT fixed point:  $0<\hat w_*^{\rm RT}<1$, always UV stable

\nd  when it exists, {\it i.e.} for $d>2$.

\end{itemize}
The complete flow diagram as a function of the dimension is displayed in Fig.~\ref{fig:flowdiag}.

The values of the critical exponents are trivially determined for the EW fixed point in all $d$ because  $\eta_*^\nu=\eta_*^{\cal D}=0$ for  $\hat w_*^{\rm EW}=0$ (see \eqref{eq:etanuD}) and thus $z=2$ and $\chi=(2-d)/2$ from \eqref{eq:defexpo}. They are also determined for the KPZ fixed point in $d=1$ since the accidental time reversal symmetry imposes $\eta_*^\nu=\eta_*^{\cal D}\equiv \eta_*$ \cite{Canet2010}, which yields $\eta_*=1/2$, and thus $z=3/2$ and $\chi=1/2$. In all other cases, the values of $\eta_*^\nu$ and $\eta_*^{\cal D}$ are to be calculated from the numerical solution of the full equations \eqref{eq:dsw}, \eqref{eq:etanuD}. At the simple level of approximation presented in this section, the estimates of the exponents are poor, see Appendix~\ref{app:IBfixed-point}. Obtaining accurate estimates requires much richer approximations \cite{Canet2010,Canet2011kpz,Kloss2012}.
However, it is worth emphasizing that the simple approximation used here already encompasses the two strong-coupling KPZ and IB fixed points in all dimensions. In contrast, the pertubative RG equations, even resummed to all orders of perturbation theory,  only contain the EW and RT fixed points \cite{Wiese98}. The perturbative RG flow just diverges for $d\geq 2$ at a finite RG scale (which corresponds to an IR Landau pole) and completely  fails to capture the strong-coupling  fixed points.

 We show in the next section that the dynamical exponent at the IB fixed point can be determined  using another approach to solve the FRG equations, which yields the exact result $z=1$.

\section{Exact flow equations in the large-momentum limit}
\label{sec:largep}

\subsection{Relevance of the IB fixed point}

We showed in the previous section, within a simple approximation, 
that the IB fixed point exists in all dimensions. 
We expect that the general fixed point structure displayed in Fig.~\ref{fig:flowdiag} does not depend on a specific approximation and remains qualitatively unchanged beyond the approximation \ref{eq:anzLPA}, and presumably in the exact theory. Thus, in the following, we assume in particular that the IB fixed point exists independently of the approximation used. This fixed point is always UV stable (IR unstable), which implies that it controls UV (large momentum) modes.  Let us explain why the associated behavior is observable in the inviscid limit.
\begin{figure}
\includegraphics[width=8cm]{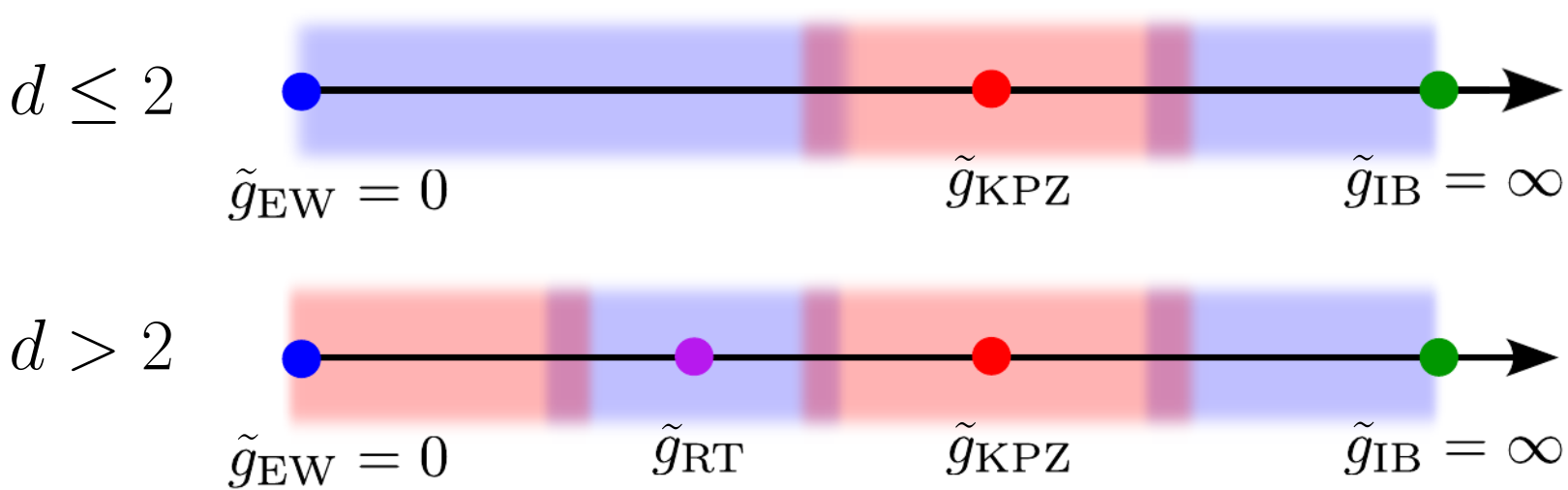}
 \caption{Sketch of the fixed-point structure of the Burgers-KPZ equation in $d\leq 2$ and in $d>2$. The fixed points are represented by the dots, the red-shaded area represents the region which controls the IR modes, while the blue-shaded area represents the region which controls the UV modes. The crossover regions correspond to scales $\kappa$ around $\kappa_{\rm IR}$  introduced in the text.}
 \label{fig:schemg}
\end{figure}

The fixed point structure of the Burgers-KPZ equation is schematically depicted in Fig.~\ref{fig:schemg}.
The RG flow tends in the IR to the  KPZ fixed point, for any initial condition $g_\Lambda$
 in $d\leq 2$, or for any  $g_\Lambda>g^{c}_\Lambda$ in $d>2$. 
  Within the simple approximation \eqref{eq:anzLPA}, one simply has $g^{c}_\Lambda=\Lambda^{2-d}\hat g_*^{\rm RT}$, but this does not hold in general when the effective average action is not constrained to keep the  form of the bare action.
   In any dimension, the KPZ fixed point $\hat g_*^{\rm KPZ}$ is fully (or locally) IR attractive.
  Let us define $g_*\equiv \hat g_*^{\rm KPZ} \Lambda^{2-d}$, which corresponds to the value of the microscopic coupling $g_\Lambda$ for which the system would already be at the IR fixed point, and thus all modes $p$ are controlled by this fixed point \footnote{Again,  the statement that the system is already at the KPZ fixed point for $g_\Lambda=g_*$ only holds within the simplest approximation \eqref{eq:anzLPA} where $\Gamma_\kappa$ is restricted to the form of the bare action. In general, the fixed point effective action is more complicated, and $g_*$ is defined as the initial value $g_\Lambda$ which minimizes the RG time to reach the IR fixed point.}.
If  $g_\Lambda\neq g_*$, then $\hat g_\kappa$ renormalizes, and one can consider that it reaches the vicinity of the IR fixed point for a RG time $s_{\rm IR}$ such that $\hat g_{s_{\rm IR}} = \alpha\hat g_*$ (if $g_\Lambda> g_*$) or $\hat g_{s_{\rm IR}} = \hat g_*/\alpha$ (if $g_\Lambda< g_*$) where $\alpha>1$ is of order one, {\it e.g.} $\alpha=2$.
 The larger $|g_\Lambda- g_*|$ is, the longer the time $|s_{\rm IR}|$, or equivalently the smaller $\kappa_{\rm IR}=\Lambda e^{s_{\rm IR}}$.
  If a mode $p$ is such that $p\lesssim \kappa_{\rm IR}$, the corresponding $\hat g_{\kappa=p}$   is  close to $\hat g_*$ and this mode is controlled by the IR fixed point. This corresponds to the red-shaded area in Fig.~\ref{fig:schemg}. In contrast, the large momentum modes $p\gtrsim \kappa_{\rm IR}$ are controlled by the associated UV fixed point, which is the IB fixed point for any $g_\Lambda > g_*$ (and otherwise the EW ($d\leq 2$) or the RT ($d>2$) fixed point).
  In other words, the larger $|g_\Lambda- g_*|$ is, the smaller the RG scale $\kappa_{\rm IR}$ which delineates IR and UV modes, and thus the larger the extent of the UV modes exhibiting the IB scaling. This situation is favored by the  limit $\nu\to 0$, since $g_\Lambda=\lambda^2 {\cal D}/\nu^3$.  This explains why this scaling arises ({\it i.e.} is observable over a broader range) in the inviscid limit.
  We now show that the corresponding dynamical scaling exponent is $z=1$, by focusing on the large $p$ behavior of the FRG flow equations.
The derivation presented in the next section does not depend on the precise form of the Gaussian noise/forcing in the Burgers action, so it applies for both the Burgers-KPZ and Burgers equations.

\subsection{Closure at large momentum for the Burgers equation.}
\label{subsec:closurelargep}

In this section, we aim at calculating the dynamical exponent at the IB fixed point.
As explained in the previous section, this amounts to studying the large momentum $|\vp|\gtrsim \kappa_{\rm IR}$ behavior of the flow equations, where $\kappa_{\rm IR}\to 0$ as $\nu\to 0$.
An approximation scheme~\cite{Blaizot2006,Blaizot2007,Benitez2012} turns out to be particularly appropriate to study this large-momentum limit. It consists in an expansion of the vertices entering a given flow equation in powers of the internal loop momentum.
This expansion is fully justified by the presence of the (scale derivative of the) regulator  ${\partial}_s \mathcal{R}_{\kappa}(\vq)$ in the loop integral. Indeed, the general property 3- (Sec.~\ref{sec:wetterich}) of the regulator implies that the integral over the internal momentum $\vq$ is cut to values
$|\vq|\lesssim \kappa$. Thus, in the limit of large external momentum $|\vp|\gg\kappa$,
 one has $|\vq|\ll |\vp|$. Because analyticity of the vertices is guaranteed at all scale $\kappa$ by construction~\cite{Dupuis2021}, one can Taylor expand them in powers of $\vq$, since $|\vq|$ is compared to $|\vp|$ and $|\vq|/|\vp|\to 0$. This expansion becomes exact in the limit of infinite $\vp$~\cite{Blaizot2006,Blaizot2007,Benitez2012}. Let us  stress that this analysis holds at or in the vicinity of a fixed point of the FRG flow equations. This ensures that, in the large-$|\vp|$ limit, only the ratio $|\vq|/|\vp|\to 0$ can appear, and not independent occurences of $|\vq|/\kappa$.

 It was realized that, in the context of turbulence, this approximation scheme leads to exact results for the space-time dependence of generic $n$-point correlation functions in the limit of large momentum. Their derivation is grounded in the existence of extended symmetries and the related Ward identities, which allows  one to close the corresponding flow equations in the large-momentum limit.
 This closure was first achieved for the incompressible Navier-Stokes equation in three, and then two dimensions~\cite{Canet2016,Canet2017,Tarpin2018,Tarpin2019,Pagani2021}, and it was recently extended for the one-dimensional Burgers-KPZ equation~\cite{Fontaine2023InvBurgers}. We now show that it can be generalized to the $d$-dimensional Burgers equation as well.

The derivation is strictly analogous to Refs.~\cite{Canet2016,Fontaine2023InvBurgers}, with the main difference that the transverse projectors (for Navier-Stokes equation) are replaced with longitudinal ones (for Burgers equation). We report this derivation for the two-point functions in Appendix~\ref{app:largep} for completeness. We emphasize that it can also be applied to arbitrary $n$-point correlation function,  following the same proof as in Ref.~\cite{Tarpin2018}. We here only discuss the result for  the two-point  connected correlation function of the velocity $C_{\kappa} (\varpi,\vp)\equiv G^{(2,0)}_{\kappa,\parallel} (\varpi,\vp)$, whose flow equation is derived in Appendix~\ref{app:largep} and  given by \eqref{eq:flowCapp}. Fourier transforming this equation back in time delay, one obtains
\begin{equation}\label{flowC}
    \partial_{\kappa} C_{\kappa} (t,\vp) =
    \frac{1}{d} \; p^2  C_{\kappa} (t,\vp)\int_{\omega}\dfrac{\cos(\omega t) -1}{\omega^2}\;\tilde{\p}_s \int_{\vq}C_{\kappa} (\omega,\vq)\,.
\end{equation}
Note that this flow equation is identical to the one obtained for the Navier-Stokes equation, apart from the $1/d$ pre-factor which arises because the longitudinal rather than the transverse part of the correlation function matters. In fact,  Appendix~\ref{app:largep} shows that the only ingredients entering the derivation of this equation are the Ward identities associated with the response-shift and Galilean time-gauged symmetries. Thus, any system sharing these basic symmetries will display similar flow equations, leading in turn to a $z=1$ dynamical exponent at large momentum (as shown below). This $z=1$ is thus purely the imprint of the Galilean and shift symmetries in a scaling regime.

\subsection{Solution at the fixed point}
\label{subsec:sollargep}

We showed  in Sec.~\ref{sec:LPA}  that the FRG flow always reaches an IR fixed point. We are interested in the inviscid limit, or equivalently the strong-coupling regime, in which case the IR fixed point is always the KPZ one in any $d$. At this fixed point,
 the (dimensionless) term $\tilde{\p}_s \int_{\hat\vq}\hat{C}_{\kappa} (\hat{\omega},\hat{\vq})$ becomes a simple $\kappa$-independent function $\hat J(\hat\omega)$. It turns out that
 the (non-linear) flow equation~\eqref{flowC} can be solved analytically in both limits of  large and small time-delays $t$ where the remaining integral over $\omega$ can be simplified (we refer the interested reader to \cite{Tarpin2018,Fontaine2023InvBurgers} for the details of the calculation).
 The solution reads as  
\begin{equation}\label{solutionC}
    C(t,\vp) =  C(0,\vp){\times}
    \begin{cases}
        \exp\left( - \mu_0 (pt)^2 \right),\quad t\ll \tau_c
        \\
        \exp\left( - \mu_{\infty} p^2 |t| \right),\quad t\gg \tau_c
    \end{cases}
\end{equation}
where $\mu_0$, $\mu_{\infty}$ are non-universal constants and $\tau_c$ is a characteristic time scale.
The computation of these constants (related to the fixed-point value of $\hat J(\hat\omega)$), or of the normalization $C(0,\vp)$, would require to integrate the whole RG flow. They are not given by the sole large momentum expansion. However, the behavior in $t$ and $p$ in the exponential argument in \eqref{solutionC} is exact in the limit of large $p$. It is of order $p^2$, while corrections (not indicated here) are at most of order $p$.
This solution \eqref{solutionC} shows that
the short-time behavior of the correlation function is a scaling form with  the variable $p t^{1/z}\equiv pt$,
 which proves that the associated dynamical exponent is $z=1$.
  The large momentum and short time regime of the Burgers equation,  thus exhibits
   the $z=1$ dynamical exponent, independently of the dimension $d$, and independently of the range of the forcing, as long as it is Gaussian and white in time. This regime arises for small $\nu$, and it extends over a broader range as $\nu\to 0$.

  The solution~\eqref{solutionC} also predicts, in the same $\nu\to 0$ limit, a crossover to another regime, with $z=2$, at large time delays. This crossover was also predicted for the Navier-Stokes equation and for the one-dimensional Burgers equation. It turns out that it is challenging to observe  it in numerical simulations, since the short-time Gaussian decay rapidly leads to noise-level  signal, which prevents from resolving the long-time behavior~\cite{Gorbunova2021}. A hint of this crossover appears in the simulation of the one-dimensional Burgers-KPZ equation of Ref.~\cite{Brachet2022} as a lack of collapse of the data at large time when the scaling variable $pt$ is used. There is one case, reported in Ref.~\cite{Gorbunova2021scalar}, where  the crossover to the large-time regime has been unambiguously evidenced, which is the case of passive scalars transported by a turbulent Gaussian velocity field with finite time correlations (generalization of the Kraichnan model).
 We expect that this large-time regime is also difficult to observe in simulations of the Burgers equation in $d>1$. In contrast, the small-time regime should be easily accessible in simulations in $d>1$.

\section{Conclusion and outlook}

In this work, we have proposed an action for the $d$-dimensional Burgers equation which properly enforces the irrotationality of the flow, while preventing  the appearance of unphysical zero modes, thanks to the introduction of auxiliary response fields. We have shown that this action possesses extended symmetries which play a similar role to the ones of the Navier-Stokes action. In the latter case, these symmetries were exploited to achieve the exact closure of the flow equations for any $n$-point correlation functions in the limit of large momentum. We have thus transposed this closure to the $d$-dimensional Burgers equation. The resulting flow equation for the two-point correlation function of the velocity can be solved exactly at the fixed point and the solution shows that it features at short time-delays a scaling regime with the exact dynamical exponent $z=1$. This scaling is deeply rooted in the Galilean and shift extended symmetries of the system, together with the existence of a non-Gaussian fixed point of the FRG equations. As a consequence, it is super-universal, in the sense that it does not depend on the dimension $d$, and it does not depend either on the scale of the stirring force/noise. It   thus fully bridges the result of~\cite{Bouchaud1995} in $d\to \infty$ with large-scale forcing and of~\cite{Fontaine2023InvBurgers} in $d=1$ with small-scale noise.
 
 The $z=1$ scaling describes the temporal behavior at  large momentum of the velocity correlation function. It gives no information on the equal-time statistics, in particular  on the structure functions, since the solution~\eqref{solutionC} becomes trivial at equal time delays $t=0$. In contrast with the large momentum $z=1$ regime, the equal-time properties are sensitive to both the dimension and the scale of the forcing. For a large-scale forcing, the multi-dimensional Burgers equation leads to turbulence with strong intermittency, and it would be very interesting to calculate the corresponding structure functions using FRG. For a small-scale noise, the behavior of the system is controlled by the KPZ fixed point, and calculating with great precision the associated universal scaling functions in all $d$ would also be very desirable and is work in progress.

\begin{acknowledgments}
L.C. acknowledges support from IUF (Institut Universitaire de France).
L.G. acknowledges support by the MSCA Cofund QuanG (Grant Number : 101081458) funded by the European Union. Views and opinions expressed are however those of the authors only and do not necessarily reflect those of the European Union or Université Grenoble Alpes. Neither the European Union nor the granting authority can be held responsible for them. N.W. thanks the `Programa de Desarrollo de las Ciencias B\'asicas' for its support and  the LPMMC for a scientific invitation during which part of the present work was done. This work received the support of the French-Uruguayan Institute of Physics (IFU$\Phi$).
\end{acknowledgments}

\appendix
\begin{widetext}

\section{Burgers action for arbitrary dimensions}
\label{app:higherd}

In this appendix, the construction of the action presented in Eq.~(\ref{eq:action}) is generalized for arbitrary dimensions. As mentioned in Sec.~\ref{sec:action}, in general dimension, the rotational of a vector does not correspond to a (pseudo-)vector, but must be viewed as a rank-two antisymmetric tensor. This leads to the fact that the action given by Eq.~(\ref{eq:action}) must be slightly modified. As shown here, once these modifications are implemented, the results of the three-dimensional case generalize to arbitrary $d$. In particular, once the equations of motion of the auxiliary fields (necessary to impose the irrotationality of the velocity and response velocity fields) are imposed, the velocity and response velocity sector is decoupled from the rest, the only imprint resulting from these constraints being  a longitudinal projector in the reduced propagator.

Firstly, the case $d=1$ has already been discussed in Sec.~\ref{sec:action}. The case $d=2$ is a trivial particular case of the $d=3$ one since in $d=2$, the $w_{\beta}$ and $\tom_{\beta}$ fields only have non-zero components with $\beta=3$, and obviously only depend on the $x_1$ and $x_2$ coordinates, so the divergence condition becomes trivial. That is, the action given by Eq.~(\ref{eq:action}) also applies to the case $d=2$ except for the fact that the terms $\propto \theta$ or $\propto \ttheta$ can be omitted.

We proceed to analyze the nontrivial case corresponding to $d>3$. In such a case, the term for imposing the irrotationality of the velocity field must be replaced by (and similarly for the response velocity field):
\begin{equation}
\tvom \cdot (\nabla \times \vv) \to \frac 1 2 \bar{w}_{\alpha\beta}(\partial_\alpha v_\beta-\partial_\beta v_\alpha)=\bar{w}_{\alpha\beta}\partial_\alpha v_\beta.
\end{equation}
As in the three-dimensional case, when $d>3$, it is necessary to cancel the longitudinal components of the field $\tvom$. In $d=3$, this is achieved by introducing a (pseudo-)scalar Lagrange multiplier $\theta$. For generic dimension, it is necessary to introduce an antisymmetric tensor of rank $d-3$, thereby including in the density of the action a term of type
\begin{equation}
 \bar{w}_{\alpha\beta} \epsilon_{\alpha\beta\gamma\mu_1\mu_2\dots} \partial_\gamma\theta_{\mu_1\mu_2\dots}\,
\end{equation}
where $\epsilon_{\alpha\beta\gamma\mu_1\mu_2\dots}$ is the completely antisymmetric pseudo-tensor of rank $d$ existing in dimension $d$. This expression naturally reduces to the term included in~(\ref{eq:action}) in $d=3$ but is valid in generic dimension. One proceeds similarly for the $\vom$ field.

The problem that arises at this point is that the included terms do not involve the longitudinal components of the ${\bf \theta}$ and ${\bf \bar{\theta}}$ fields which, in constrast with the three-dimensional case,  appear for higher dimensions. The solution to this difficulty is to iterate the procedure of imposing the cancellation of the longitudinal components by new Lagrange multipliers in the same way as described for the $\vom$ and $\tvom$ fields. For example, in dimension $d=4$ the ${\bf \theta}$ and ${\bf \bar{\theta}}$ fields are vectors. In such a case it is sufficient to add to the density of the action a term of type
\begin{equation}
 \bar \chi \partial_\alpha \theta_\alpha
\end{equation}
(and a similar one for $\bar \theta$). In general dimension $d$, this iterative procedure stops after adding $d-1$ pairs of Lagrange multipliers, the last pair being scalars.

 Since all the added auxiliary fields enter through quadratic terms in the action, they lead to similar extended symmetries (fully gauged shifts) as in $d=3$.
 It follows that the auxiliary field sector is not renormalized, or otherwise stated that the effective action  $\Gamma$ keeps the same form as the bare action $\mathcal{S}$ in this sector. Moreover,  we have checked that in any dimension, this procedure respects the extended response-fields shift and Galilean symmetries, and that the Hessian of the action is invertible, resulting in a well-defined propagator. Finally, once the equations of motion of the auxiliary fields are employed,  one can restrict to the reduced velocity and response velocity sector.  The only remnant of the irrotationality constraints is the longitudinal projector in the  reduced propagator.

\section{General structure of the propagator matrix}
\label{app:prop}

In this section,  the derivation is presented in the more intuitive $d=3$ notation, but it can be generalized for arbitrary $d$ following the procedure described in Appendix \ref{app:higherd}. Most importantly, one obtains that the general form of the reduced propagator matrix in the $(\vu,\tvu)$ sector is given by \eqref{eq:G11}, \eqref{eq:G20} in any $d>1$.

The general form of the Hessian of the effective average action $\Gamma_\kappa$ can be inferred by taking two functional derivatives of $\Gamma_\kappa$, whose  general structure is given by~\eqref{eq:Gammakappa}.
Transforming to Fourier space, one obtains
\begin{equation}
\bar\Gamma^{(2)}_{\kappa}(\omega,\vp) =
\begin{blockarray}{ccccccc}
    \quad\quad\quad\quad       & u_\beta & \tu_\beta &  w_\beta  &  \quad\tom_\beta  &   \theta  &  \ttheta \\
\begin{block}{c(cccccc)}
 	u_\alpha      & 0 & \bar\Gamma^{(1,1)}_{\kappa,\alpha\beta}(-\omega,\vp) & 0 & \;\;i \epsilon_{\alpha\gamma\beta} p_{\gamma}\;\; & 0 & 0 \\
	\tu_\alpha & \bar\Gamma^{(1,1)}_{\kappa,\alpha\beta}(\omega,\vp)   & \bar\Gamma^{(0,2)}_{\kappa,\alpha\beta}(\omega,\vp)  & i \epsilon_{\alpha\gamma\beta} p_{\gamma} & 0 & 0 & 0  \\
	w_\alpha & 0 & i \epsilon_{\alpha\gamma\beta} p_{\gamma} & 0  & 0 & 0 & \;\;\;\;-i p_\alpha\;\;\;\; \\
	\tom_\alpha      & \;\;i \epsilon_{\alpha\gamma\beta} p_{\gamma}\;\; & 0 & 0 & 0 & \;\;\;\;i p_\alpha\;\;\;\; & 0 \\
	\theta   & 0 & 0 & 0 & -i p_\beta & 0 & 0 \\
	\ttheta  & 0 & 0 & i p_\beta & 0 & 0 & 0 \\
\end{block}
\end{blockarray}
\end{equation}
using the general notation \eqref{eq:vertexmn} and \eqref{eq:vertexFourier} for the vertex functions.
The regulator matrix is chosen as
\begin{equation}
{\cal R}_{\kappa}(\omega,\vp) =
\begin{blockarray}{ccccccc}
    \quad\quad\quad\quad       & u_\beta & \tu_\beta & w_\beta & \tom_\beta & \theta &  \ttheta \\
\begin{block}{c(cccccc)}
 	u_\alpha      &\;\;\;\;0\;\;\;\; & \;\;{\cal M}_{\kappa}(\vp)\;\; & \;\;\;\;0 \;\;\;\;&\;\;\;\;0\;\;\;\; &\;\;\;\;0 \;\;\;\;& \;\;\;\;0 \;\;\;\;\\
	\tu_\alpha &  {\cal M}_{\kappa}(\vp)  & {\cal N}_{\kappa}(\vp)  & 0 & 0 & 0 & 0  \\
	w_\alpha & 0 & 0 & 0  & 0 & 0 & 0 \\
	\tom_\alpha      &0 & 0 & 0 & 0 & 0 & 0 \\
	\theta   & 0 & 0 & 0 & 0 & 0 & 0 \\
	\ttheta  & 0 & 0 & 0 & 0 & 0 & 0 \\
\end{block}
\end{blockarray} \quad P^\parallel_{\alpha\beta}(\vp)
\end{equation}
The propagator is defined as the inverse of the matrix $\bar\Gamma_\kappa^{(2)}+{\cal R}_\kappa$, which is given by
\begin{equation}
\bar{\cal G}_{\kappa}(\omega,\vp) =
\begin{blockarray}{ccccccc}
      \quad\quad\quad\quad        & u_\beta & \tu_\beta & w_\beta & \tom_\beta & \theta &  \ttheta \\
\begin{block}{c(cccccc)}
	u_\alpha     & \bar{G}^{(2,0)}_{\kappa,\alpha\beta}(\omega,\vp) & \bar{G}^{(1,1)}_{\kappa,\alpha\beta}(\omega,\vp) & 0 & \dfrac{i \epsilon_{\alpha\gamma\beta} p_{\gamma}}{p^2} & 0 & 0 \\
	\tu_\alpha & \bar{G}^{(1,1)}_{\kappa,\alpha\beta}(-\omega,\vp)    & 0 & \dfrac{i \epsilon_{\alpha\gamma\beta} p_{\gamma}}{p^2} & 0 & 0 & 0  \\
	w_\alpha & 0 & \dfrac{i \epsilon_{\alpha\gamma\beta} p_{\gamma}}{p^2} & -\dfrac{1}{p^2}\bar\Gamma^{(0,2)}_{\kappa,\perp}(\omega,\vp) P^{\perp}_{\alpha\beta}  & -\dfrac{1}{p^2}\bar\Gamma^{(1,1)}_{\kappa,\perp}(\omega,\vp) P^{\perp}_{\alpha\beta} & 0 & \;\;\;\;-\dfrac{i p_\alpha}{p^2}\;\;\;\; \\
	 \tom_\alpha      & \dfrac{i \epsilon_{\alpha\gamma\beta} p_{\gamma}}{p^2} & 0 & -\dfrac{1}{p^2}\bar\Gamma^{(1,1)}_{\kappa,\perp}(-\omega,\vp) P^{\perp}_{\alpha\beta}  & 0 & \;\;\;\;\dfrac{i p_\alpha}{p^2}\;\;\;\; & 0 \\
	\theta 		& 0 & 0 & 0 &-\dfrac{i p_\beta}{p^2} & 0 & 0 \\
	\ttheta  & 0 & 0 & \dfrac{i p_\beta}{p^2} & 0 & 0 & 0 \\
\end{block}
\end{blockarray}
\end{equation}
where
\begin{align}
   \bar{G}^{(1,1)}_{\kappa,\alpha\beta}(\omega,\vp)  &\equiv
    \dfrac
    {1}
    {
     \bar\Gamma^{(1,1)}_{\kappa,\parallel}(\omega,\vp) + {\cal M}_{\kappa}(\vp)
    } 
     P^{\parallel}_{\alpha\beta}(\vp), \label{eq:G11}\\
    \bar{G}^{(2,0)}_{\kappa,\alpha\beta}(\omega,\vp)  &\equiv
    \dfrac
    {-(\bar\Gamma^{(0,2)}_{\kappa,\parallel}(\omega,\vp) + {\cal N}_{\kappa}(\vp))}
    {\left|
    \bar\Gamma^{(1,1)}_{\kappa,\parallel}(\omega,\vp) + {\cal M}_{\kappa}(\vp)
    \right|^2} 
     P^{\parallel}_{\alpha\beta}(\vp)\,, \label{eq:G20}
\end{align}
with the longitudinal and transverse projectors defined as
\begin{equation}
  P^{\parallel}_{\alpha\beta}(\vp) = \frac{p_{\alpha} p_{\beta}}{p^2}\,, \qquad  P^{\perp}_{\alpha\beta}(\vp) = \delta_{\alpha\beta} - \frac{p_{\alpha} p_{\beta}}{p^2} \,.
\end{equation}

\section{Flow equations within the  approximation \eqref{eq:anzLPA}}
\label{app:IBfixed-point}

The flow equations for $\nu_\kappa$ and ${\cal D}_\kappa$ can be obtained 
following the definitions \eqref{eq:defNu} from the flows of 
$\Gamma_{\kappa,\parallel}^{(1,1)}$ and $\Gamma_{\kappa,\parallel}^{(0,2)}$, given by the exact equations \eqref{eq:dsgam2}. 
 Within the simple approximation corresponding to the ansatz~\eqref{eq:anzLPA}, only the vertex $\Gamma_{\kappa}^{(2,1)}$ is non-zero, such that only the diagrams $(c)$ of Fig.~\ref{fig:flowG11} and \ref{fig:flowG02} contribute.
 They are given by
 \begin{align}
 \left[\partial_s \bar \Gamma_{\kappa,\parallel}^{(1,1)}(\varpi,   \vp)\right]_{(c)}
 =&-  \tilde\partial_s\Bigg[ P^{\parallel}_{\alpha\beta}(\vp)  \int_{\omega,  \vq} \bar \Gamma_{\kappa,\alpha ij }^{(2,1)}(\varpi,
\vp,\omega,  \vq)\bar G^{(1,1)}_{\kappa,jk}(\omega+\varpi,  \vp+\vq)
 \bar \Gamma_{\kappa,kl\beta}^{(2,1)}(\omega+\varpi,  \vp+\vq,-\omega,  -\vq)\; 
\bar G^{(2,0)}_{\kappa,li}(\omega,  \vq) \Bigg] \\
 \left[\partial_s \bar \Gamma_{\kappa,\parallel}^{(0,2)}(\varpi,   \vp)\right]_{(c)} & =- \dfrac 1 2
 \tilde\partial_s \Bigg[ P^{\parallel}_{\alpha\beta}(\vp) \int_{\omega,\vq} \bar \Gamma_{\kappa,ij\alpha}^{(2,1)}(\omega,  \vq,-\omega-\varpi,-  \vp-\vq)\bar G^{(2,0)}_{\kappa,jk}(\omega+\varpi,  \vp+\vq)  \nonumber\\
 &
 \qquad\qquad\qquad\times\bar \Gamma_{\kappa,kl\beta}^{(2,1)}(\omega+\varpi,  
\vp+\vq,-\omega,  -\vq) \;\bar G^{(2,0)}_{\kappa,li}(\omega,  \vq)\Bigg]\, .
 \end{align}
 One then replaces the propagators $\bar G^{(2,0)}_{\kappa}$ and  $\bar G^{(1,1)}_{\kappa}$ with their explicit expressions \eqref{eq:G11} and  \eqref{eq:G20}, and inserts the simplified expressions \eqref{eq:gam2LPA}  and  \eqref{eq:gam21} for the vertex functions  $\bar\Gamma_{\kappa}^{(1,1)}$, $\bar\Gamma_{\kappa}^{(0,2)}$, and $\bar\Gamma_{\kappa}^{(2,1)}$. Applying the $\tilde\partial_s$ derivative, and  the $\partial_{p_\alpha}^2$ derivative, one obtains
\begin{align}
 \p_s \nu_\kappa &= \frac{\lambda^2}{d} \int \dfrac{d\omega}{2\pi}\dfrac{d^d\vq}{(2\pi)^d}
   \frac{1}{\big(\ell^2_\kappa(y) + \omega^2\big)^3} \times
   \nonumber\\
   &\Bigg\{2 \p_s{\cal M}_\kappa(y)
     \Big[k_\kappa(y)\big( (d-3)\ell^2_\kappa(y) -(d-1)\omega^2 \big)  + 2 y\p_y k_\kappa(y)\big(\ell^2_\kappa(y)-\omega^2\big) \Big] \nonumber\\
    & +  \p_s{\cal N}_\kappa(y) \Big[ -\ell_\kappa(y)\big(\ell^2_\kappa(y)+\omega^2\big)+2 y\p_y \ell_\kappa(y) \big(\ell^2_\kappa(y)-\omega^2\big)\Big] \Bigg\}
 \, , \\
 \p_s {\cal D}_\kappa &= - {\lambda^2} \int \dfrac{d\omega}{2\pi}\dfrac{d^d\vq}{(2\pi)^d} \dfrac{k_\kappa(y)}{\big(\ell_\kappa^2(y) + \omega^2\big)^3}\;\Big[
   4 k_\kappa(y) \ell_\kappa(y) \p_s{\cal M}_\kappa(y) + \big(\ell_\kappa^2(y) + \omega^2\big)\p_s{\cal N}_\kappa(y)\Big] \, ,
\end{align}
where $y\equiv q^2$, $\ell_\kappa(y) \equiv y \nu_\kappa + {\cal M}_\kappa(y)$, 
$k_\kappa(y) \equiv y{\cal D}_\kappa - \frac{1}{2}{\cal N}_\kappa(y)$, and with $\lambda_\kappa\equiv \lambda$ and $\mu_\kappa\equiv 1$  since they are not renormalized.
 We assumed the cut-off functions ${\cal M}_\kappa$, ${\cal N}_\kappa$ to depend only on  $y=q^2$, and we omitted terms which are odd in $\omega$ since they vanish upon integration.
The frequency integrals can be explicitly carried out, yielding
\begin{align}
 \p_s \nu_\kappa &= \frac{\lambda^2}{4d} \int \dfrac{d^d\vq}{(2\pi)^d} \dfrac{1}{\ell_\kappa^3(y)}
   \Big[\Big( (d-4) k_\kappa(y) +2 y \p_{y}k_\kappa(y) \Big) \p_s{\cal M}_\kappa(y) +
   \Big( - {\ell}_\kappa(y) + y\p_{y}\ell_\kappa(y) \Big) \p_s{\cal N}_\kappa(y)
 \Big]
 \, ,\\
 \p_s {\cal D}_\kappa &= - \frac{\lambda^2}{4} \int \dfrac{d^d\vq}{(2\pi)^d} \dfrac{ k_\kappa(y)}{\ell_\kappa^4(y)}\Big[
   3 k_\kappa(y) \p_s{\cal M}_\kappa(y) + \ell_\kappa(y) \p_s{\cal N}_\kappa(y)
 \Big]
 \,.
\end{align}
We then introduce dimensionless variables according to \eqref{eq:rescaling}, and write the cut-off functions
in the form ${\cal M}_\kappa(y) = \nu_\kappa y \,\hat{m}(\hat y)$ and ${\cal
N}_\kappa(y) = -2 {\cal D}_\kappa \,y\,\hat{n}(\hat y)$ with $\hat y=y/\kappa^2$, such that one has
\begin{align}
 \p_s {\cal M}_\kappa(y) &= -\nu_\kappa y \big(\eta_\kappa^\nu \hat{m}(\hat y) + 2 \hat y\,\hat{m}'(\hat y)\big)\equiv -\nu_\kappa y \;s_{\cal M}(\hat y) \, \\
 \p_s {\cal N}_\kappa(y) &= 2{\cal D}_\kappa {y} \big(\eta_\kappa^{\cal D} \hat{n}(\hat y) + 2 \hat y\,\hat{n}'(\hat y)\big)\equiv 2{\cal D}_\kappa  {y}\, s_{\cal N}(\hat y)\,.
\end{align}
The dimensionless flows of the running anomalous dimensions,
defined as $ \eta_\kappa^\nu = -\partial_s \ln \nu_\kappa$ and $ \eta_\kappa^{\cal D} = -\partial_s \ln {\cal D}_\kappa$, are given by
\begin{align}
 \eta_\kappa^\nu & =
 \hat g_\kappa \frac{v_d}{8d} \int_0^\infty d\hat{y} \dfrac{\hat{y}^{d/2-2}}{
   (1+\hat{m}(\hat{y}))^3
 }
 \Big[
   s_{\cal M}(\hat{y})\Big((d-2)(1+\hat{n}(\hat{y})) + 2\hat{y}\hat{n}'(\hat{y})\Big)
   -
   2 s_{\cal N}(\hat{y})\,\hat{y}\hat{m}'(\hat{y})
 \Big]
\,, \\
 \eta_\kappa^{\cal D} & = -
 \hat g_\kappa \frac{v_d}{8} \int_0^\infty d\hat{y} \dfrac{\hat{y}^{d/2-2}}{\big(1+\hat{m}(\hat{y})\big)^4}
 \Big[
   (1+\hat{n}(\hat{y})) \Big(
     3 (1+\hat{n}(\hat{y})) s_{\cal M}(\hat y)  -
     2 (1+\hat{m}(\hat{y})) s_{\cal N}(\hat y) \Big)
   \Big] \,,
\end{align}
with $\hat{n}'(\hat{y})\equiv \p_{\hat{y}}\hat{n}(\hat{y})$ and $\hat{m}'(\hat{y})\equiv \p_{\hat{y}}\hat{m}(\hat{y})$, and where we defined $v_d = ( 2^{d-1}\pi^{d/2}\Gamma(d/2))^{-1}$.
Solving this linear system for $\eta_\kappa^\nu$ and $\eta_\kappa^{\cal D}$ then yields
\begin{align}
 \eta_\kappa^\nu &= \frac{
   {\cal I}^\nu_0 - {\cal I}^{\cal D}_{\cal D} {\cal I}^\nu_0 + {\cal I}^{\cal D}_0 {\cal I}^\nu_{\cal D}
 }{
   1 - {\cal I}^{\cal D}_{\cal D} - {\cal I}^\nu_\nu - {\cal I}^{\cal D}_\nu {\cal I}^\nu_{\cal D} + {\cal I}^{\cal D}_{\cal D} {\cal I}^\nu_\nu
 }
 \,,  \\
 \eta_\kappa^{\cal D} &= \frac{
   {\cal I}^{\cal D}_0 - {\cal I}^\nu_\nu {\cal I}^{\cal D}_0 + {\cal 
   I}^\nu_0 {\cal I}^{\cal D}_\nu
 }{
   1 - {\cal I}^{\cal D}_{\cal D} - {\cal I}^\nu_\nu - {\cal I}^{\cal D}_\nu {\cal I}^\nu_{\cal D} + {\cal I}^{\cal D}_{\cal D} {\cal I}^\nu_\nu
 } \,,
\label{eq:etanuD}
\end{align}
where
\begin{align}
 {\cal I}^\nu_0 &\equiv \hat g_\kappa \frac{v_d}{4d} \int_0^\infty d\hat{y} \hat{y}^{d/2-1} \frac{
   (d-2)(1+\hat{n}(\hat{y}))\hat{m}'(\hat{y})
 }{
   (1+\hat{m}(\hat{y}))^3
 }
\,,\\
 {\cal I}^\nu_\nu &\equiv \hat g_\kappa \frac{v_d}{8d} \int_0^\infty d\hat{y} \hat{y}^{d/2-2} \frac{
   \left[(d-2)(1+\hat{n}(\hat{y})) + 2\hat{y}\hat{n}'(\hat{y})\right]\hat{m}(\hat{y})
 }{
   (1+\hat{m}(\hat{y}))^3
 }
\,,\\
 {\cal I}^\nu_{\cal D} &\equiv  -\hat g_\kappa \frac{v_d}{4d} \int_0^\infty d\hat{y} \hat{y}^{d/2-1} \frac{
   \hat{n}(\hat{y})\hat{m}'(\hat{y})
 }{
   (1+\hat{m}(\hat{y}))^3
 }
\,,\\
 {\cal I}^{\cal D}_0 &\equiv - \hat g_\kappa \frac{v_d}{4} \int_0^\infty d\hat{y} \hat{y}^{d/2-1} \frac{
   (1+\hat{n}(\hat{y}))}{
   (1+\hat{m}(\hat{y}))^4
 }\Big[3(1+\hat{n}(\hat{y}))\hat{m}'(\hat{y}) -2
   (1+\hat{m}(\hat{y}))\hat{n}'(\hat{y})\Big]
\,,\\
 {\cal I}^{\cal D}_\nu &\equiv - \hat g_\kappa \frac{3v_d}{8}
 \int_0^\infty d\hat{y} \hat{y}^{d/2-2} \frac{
   (1+\hat{n}(\hat{y}))^2\hat{m}(\hat{y})
 }{
   (1+\hat{m}(\hat{y}))^4
 }
\,,\\
 {\cal I}^{\cal D}_{\cal D} &\equiv \hat g_\kappa \frac{v_d}{4}
 \int_0^\infty d\hat{y} \hat{y}^{d/2-2} \frac{
   (1+\hat{n}(\hat{y}))\hat{n}(\hat{y})
 }{
   (1+\hat{m}(\hat{y}))^3
 }
\,.
\end{align}
It is clear that the  integrals ${\cal I}^\nu_0$, ${\cal I}^\nu_\nu$, ${\cal I}^\nu_{\cal D}$, ${\cal I}^{\cal D}_0$, ${\cal I}^{\cal D}_\nu$, ${\cal I}^{\cal D}_{\cal D}$ strongly depend on the choice of the cutoff functions $\hat{m}(y)$ and $\hat{n}(y)$.
Since with this simple approximation we are only aiming at a qualitative picture of the phase
diagram, we conveniently choose the regulator $\hat{m}(y)=\hat{n}(y)=(1/y-1)\theta(1-y)$, which allows one to analytically perform the momentum integrals. Defining $\tilde{g}_\kappa =  v_d \,\hat{g}_\kappa$ One obtains
\begin{align}
 \eta_\kappa^\nu & = - \frac{
   (d+2) \tilde{g}_\kappa \left(d^3- d (\tilde{g}_\kappa+4) + \tilde{g}_\kappa\right)
 }{
    2d^5 + 8 d^4 + d^3 (8 - 3 \tilde{g}_\kappa) - 2 d^2 \tilde{g}_\kappa +  d \tilde{g}_\kappa (\tilde{g}_\kappa + 8) - \tilde{g}_\kappa^2
 }
 \,, \\
 \eta_\kappa^{\cal D} & = \frac{
   (d+2) \tilde{g}_\kappa \left(d^3 + 2d^2 + \tilde{g}_\kappa(d-1)\right)
 }{
   2d^5 + 8 d^4 + d^3 (8 - 3 \tilde{g}_\kappa) - 2 d^2 \tilde{g}_\kappa +  d \tilde{g}_\kappa (\tilde{g}_\kappa + 8) - \tilde{g}_\kappa^2
 }
 \,.
\end{align}

Although this simple approximation provides the complete qualitative phase diagram of the Burgers-KPZ equation in all $d$, the numerical values of the exponents for $d>1$ are very poor, and strongly dependent on the choice of the regulator. As an example, in $d=2$, one obtains for this simple regulator $z\simeq 0.67$ at the KPZ fixed point (reference value from numerics being $z\simeq 1.61$ \cite{Kelling11,Oliveira13}). Deforming it to $\hat{m}(y)=\hat{n}(y)=\alpha(1/y-1)\theta(1-y)$ with $\alpha$ in the range $\alpha\in[1,10]$, the corresponding exponents vary in the range $z\in[0.67,1.20]$. For the IB fixed point, the exponent $z$ takes negative values in this whole $\alpha$-range instead of $z=1$. The dependence becomes even stronger as $d$ increases, where negative values can  also be found at both the KPZ and IB fixed points. Quantitative estimates require improved approximations, such as the ones presented in Refs.~\cite{Canet2011kpz,Kloss2012}, or alternative schemes such as the large momentum expansion implemented in this work.

\section{Calculation of the FRG flow equations in the large-momentum limit}
\label{app:largep}

We consider the exact flow equations for the two-point functions $\bar{\Gamma}_\kappa^{(1,1)}(\varpi,\vp)$ and $\bar{\Gamma}_\kappa^{(0,2)}(\varpi,\vp)$, given by~\eqref{eq:dsgam2}, and  represented diagrammatically in Fig.~\ref{fig:flowG02} and \ref{fig:flowG11}. These equations can be closed  (expressed in terms of two-point functions only) in the limit of large momentum $p=|\vp|$,   using the Ward identities.
 Indeed, as explained in  Sec.~\ref{sec:largep}, in the large-$p$ limit, one can set $\vq=0$ in all the vertices, where $\vq$ is the internal momentum, circulating in the loop. 
  If this momentum is carried by a  $\tvu$ leg (represented by an outgoing arrow), then the corresponding vertex is zero because of the Ward identity~\eqref{eq:tshiftWIvertex}, and the whole diagram vanishes. This occurs for the 
 diagrams $(b)$, $(d)$ and $(e)$ both in Fig.~\ref{fig:flowG02} and in Fig.~\ref{fig:flowG11}, which  can hence be neglected in the large-$p$ limit.

To evaluate  the remaining diagrams,  one first distribute the $\tilde \p_s$ operator. This yields either a $\tilde \p_s \bar G^{(2,0)}_{\kappa,ij}(\vQ)$ or  a $\tilde \p_s \bar G^{(1,1)}_{\kappa,ij}(\vQ)$, where $\vQ=\pm \vq$ or $\vQ=\pm (\vp+\vq)$. When $\vQ=\pm (\vp+\vq)$, one can change variables in the integrals to $\vq' = \mp (\vp+\vq)$ (and $\omega' =  \mp (\varpi+\omega))$ such that it is always the momentum $\vq$ which is cutoff in the loop. After this operation, some terms turn out to vanish again because of~\eqref{eq:tshiftWIvertex} since the $\vq$ momentum appears to be carried by a $\tvu$ leg.
For the remaining non-zero terms, the $\vq=0$ momentum is carried by a $\vu$ leg (represented by an ingoing arrow). In fact,  the corresponding diagrams always have the $\tilde\p$ operator acting on a line with a black dot ($\bar G^{(2,0)}$ line).
The vertices in these diagrams can be expressed in terms of two-point functions using the Ward identity~\eqref{eq:galileanWIvertex} related to the extended Galilean symmetry, as we detail below.

\subsection{Flow equation of $\Gamma^{(1,1)}_{\kappa,\parallel}(\varpi,\vp)$ in the  large-$\vp$ limit}

The diagram $(a)$ of Fig.~\ref{fig:flowG11} can be expressed as
\begin{align}
\left[\partial_s \bar \Gamma^{(1,1)}_{\kappa,\parallel}(\varpi, 
  p)\right]_{(a)}&=\frac 1 2  P^{\parallel}_{\alpha\beta}(\vp) 
\int_{\omega,  \vq} \bar \Gamma_{\kappa,ij\alpha\beta }^{(3,1)}(\omega, \vq,-\omega,   
-\vq, \varpi,  \vp) \;\tilde\partial_s  \bar G^{(2,0)}_{\kappa,ji}(\omega,  \vq) \nonumber \\
&\stackrel{p\to \infty}{=}\frac 1 2  P^{\parallel}_{\alpha\beta}(\vp) 
\int_{\omega,  \vq} \bar \Gamma_{\kappa,ij\alpha\beta }^{(3,1)}(\omega, 0,-\omega,   
0, \varpi,  \vp) \;\tilde\partial_s  \bar G^{(2,0)}_{\kappa,ji}(\omega,  \vq) \nonumber \\
&=\frac 1 2 \dfrac{p^2}{d} \int_\omega \frac 1 {\omega^2}\Bigg[
\bar \Gamma^{(1,1)}_{\kappa,\parallel}(\omega+\varpi,  \vp) -2 \bar \Gamma^{(1,1)}_{ \kappa,\parallel}(\varpi,  \vp) 
+\bar \Gamma^{(1,1)}_{ \kappa,\parallel}(-\omega+\varpi,  \vp)  \Bigg] \;\tilde \p_s\int_{  \vq} 
\bar G^{(2,0)}_{\kappa,\parallel}(\omega,  \vq)\, ,
\label{diag1A}
\end{align}
where the  large-$p$ limit is taken in the second line, and the identity~\eqref{eq:galileanWIvertex} is used twice to obtain the third line.
The non-zero contribution of diagram $(c)$  can be written as
\begin{align}
\left[\partial_s \bar \Gamma_{\kappa,\parallel}^{(1,1)}(\varpi,   \vp)\right]_{(c)}
 =&- P^{\parallel}_{\alpha\beta}(\vp) \int_{\omega,  \vq} \bar \Gamma_{\kappa,\alpha ij }^{(2,1)}(\varpi,
\vp,\omega,  \vq)\bar G^{(1,1)}_{\kappa,jk}(\omega+\varpi,  \vp+\vq)
 \bar \Gamma_{\kappa,kl\beta}^{(2,1)}(\omega+\varpi,  \vp+\vq,-\omega,  -\vq)\;\tilde\partial_s 
\bar G^{(2,0)}_{\kappa,li}(\omega,  \vq)\nonumber\\
&\stackrel{p\to \infty}{=} - \dfrac{p^2}{d} \int_{\omega}
\left[\frac{\bar\Gamma^{(1,1)}_{\kappa,\parallel}(\omega+\varpi,  \vp)- \bar\Gamma^{(1,1)}_{\kappa,\parallel}(\varpi, 
\vp)}{\omega}\right]^2 \bar G^{(1,1)}_{\kappa,\parallel}(\omega+\varpi, \vp)
\;\tilde\partial_s \int_{ \vq} \bar G^{(2,0)}_{\kappa,\parallel}(\omega,\vq)\,,
\label{diag1C}
\end{align}
 where the three-point vertices in the first line are evaluated in their large-$p$ limit (setting $\vq=0$) and expressed using the identity~\eqref{eq:galileanWIvertex}. Summing up the two contributions~\eqref{diag1A} and~\eqref{diag1C}, one obtains
\begin{align}
\partial_s \bar\Gamma^{(1,1)}_{\kappa,\parallel}(\varpi,  \vp)&=  \dfrac{p^2}{d}
\int_{\omega}   \Bigg\{ -\left[\frac{\bar \Gamma^{(1,1)}_{\kappa,\parallel}(\omega+\varpi, \vp)- 
\bar \Gamma^{(1,1)}_{\kappa,\parallel}(\varpi,  \vp)}{\omega}\right]^2  \bar G^{(1,1)}_{\kappa,\parallel}(\omega+\varpi, \vp) \nonumber\\
& + \frac 1 {2 \omega^2}\Bigg[\bar \Gamma^{(1,1)}_{\kappa,\parallel}(\varpi+\omega, \vp) -2 
\bar\Gamma^{(1,1)}_{\kappa,\parallel}(\varpi,  \vp) +\bar\Gamma^{(1,1)}_{\kappa,\parallel}(\varpi-\omega, \vp)  
\Bigg] \Bigg\}
\times \tilde\partial_s \int_{\vq} \bar G^{(2,0)}_{\kappa,\parallel}(\omega,\vq) \, .
\label{dtGam11}
\end{align}

\subsection{Flow equation of $\Gamma^{(0,2)}_{\kappa,\parallel}(\varpi,\vp)$ in the  large-$\vp$ limit}

Similarly, the diagram $(a)$ of Fig.~\ref{fig:flowG02} can be expressed, in the  large-$p$ limit and using  twice~\eqref{eq:galileanWIvertex}, as
\begin{align}
\left[\partial_s \bar \Gamma_{\kappa,\parallel}^{(0,2)}(\varpi,   \vp) \right]_{(a)}
&=\frac 1 2  P^{\parallel}_{\alpha\beta}(\vp) \int_{\omega} 
\bar \Gamma_{\kappa,ij\alpha\beta}^{(2,2)}(\omega, \vq,- \omega,  -\vq,\varpi, \vp)\; 
 \int_{\omega, \vq} \tilde\partial_s  \bar G^{(2,0)}_{\kappa,ji}(\omega,  \vq)\nonumber\\
&\stackrel{p\to \infty}{=} \frac 1 2 \dfrac{p^2}{d} \int_\omega \frac 1
{\omega^2}\left[ \bar \Gamma^{(0,2)}_{\kappa,\parallel}(\omega+\varpi,  \vp) -2 
\bar \Gamma^{(0,2)}_{\kappa,\parallel}(\varpi,  \vp) +\bar \Gamma^{(0,2)}_{\kappa,\parallel}(-\omega+\varpi,  \vp)  
\right]\; \tilde\partial_s \int_{  \vq} \bar G^{(2,0)}_{\kappa,\parallel}(\omega,  \vq)\, . 
\label{diag2A}
\end{align}
The expression for the diagram $(c)$ is obtained using~\eqref{eq:galileanWIvertex} for the two vertices, and reads as
\begin{align}
 \left[\partial_s \bar \Gamma_{\kappa,\parallel}^{(0,2)}(\varpi,   \vp)\right]_{(c)} & =- 
 P^{\parallel}_{\alpha\beta}(\vp) \int_{\omega,\vq} \bar \Gamma_{\kappa,ij\alpha}^{(2,1)}(\omega,  \vq,-\omega-\varpi,-  \vp-\vq)\bar G^{(2,0)}_{\kappa,jk}(\omega+\varpi,  \vp+\vq) \nonumber\\
 &
 \qquad\qquad\qquad\times\bar \Gamma_{\kappa,kl\beta}^{(2,1)}(\omega+\varpi,  
\vp+\vq,-\omega,  -\vq) \; \tilde\partial_s \bar G^{(2,0)}_{\kappa,li}(\omega,  \vq)\nonumber\\
 &\stackrel{p\to \infty}{=}  - \dfrac{p^2}{d} \int_{\omega} \frac 1
{\omega^2}\left(\bar \Gamma^{(1,1)}_{\kappa,\parallel}(-\varpi,   \vp)- 
\bar \Gamma^{(1,1)}_{\kappa,\parallel}(-\varpi-\omega,   \vp)\right)\nonumber \\ 
 &\times \left(\bar \Gamma^{(1,1)}_{\kappa,\parallel}(\varpi,   \vp)- 
\bar \Gamma^{(1,1)}_ \kappa(\varpi+\omega,   \vp)\right) 
\bar G^{(2,0)}_{\kappa,\parallel}(\omega+\varpi,  \vp) \; \tilde\partial_s \int_{  \vq} 
\bar G^{(2,0)}_{\kappa,\parallel}(\omega,  \vq)\,.
\label{diag2C}
\end{align}
Finally, the expression for 
 the diagram $(f)$ is obtained in the same way yielding
\begin{align}
\left[\partial_s \bar \Gamma_{\kappa,\parallel}^{(0,2)}(\varpi,   \vp)\right]_{(f)}
 &= - P^{\parallel}_{\alpha\beta}(\vp) \int_{\omega,  \vq} \bar \Gamma_{\kappa,ij\alpha}^{(1,2)}(\omega,  \vq,-\omega-\varpi,  -\vp-\vq)\bar G^{(1,1)}_{\kappa,jk}(\omega+\varpi,  \vp+\vq)\nonumber\\
  &
 \qquad\qquad\qquad\times
 \bar \Gamma_{\kappa,kl\beta}^{(2,1)}(\omega+\varpi,  \vp+\vq,-\omega,  -\vq)\;\tilde\partial_s
\bar G^{(2,0)}_{\kappa,li}(\omega, \vq)+c.c.\nonumber\\
&\stackrel{p\to \infty}{=}  -  \dfrac{p^2}{d} \int_{\omega}
\left[\frac{\bar \Gamma^{(0,2)}_{\kappa,\parallel}(\omega+\varpi,   \vp)- \bar \Gamma^{(0,2)}_{\kappa,\parallel}(\varpi,  \vp)}{\omega}\right] \times
\left[\frac{\bar \Gamma^{(1,1)}_{\kappa,\parallel}(\omega+\varpi,   \vp)- \bar \Gamma^{(1,1)}_{\kappa,\parallel}(\varpi, 
  \vp)}{\omega}\right] \nonumber\\
 &\qquad \qquad\times \bar G^{(1,1)}_{\kappa,\parallel}(\omega+\varpi,  \vp) \;\tilde\partial_s \int_{
\vq} \bar G^{(2,0)}_{\kappa,\parallel}(\omega,  \vq)+c.c.\, ,\label{diag2F}
\end{align}
where $c.c.$ denotes the complex conjugate.
Adding the three contributions~\eqref{diag2A},~\eqref{diag2C} and~\eqref{diag2F} finally leads to
\begin{align}
\partial_s \bar\Gamma^{(0,2)}_{\kappa,\parallel}(\varpi,  \vp)&=  \dfrac{p^2}{d}
\int_{\omega}   \Bigg\{-
 \Bigg|\frac{\bar\Gamma^{(1,1)}_{\kappa,\parallel}(\varpi,  \vp)- \bar\Gamma^{(1,1)}_{\kappa,\parallel}(\varpi+\omega,  \vp)}{\omega} \Bigg|^2 \,    \bar G^{(2,0)}_{\kappa,\parallel}(\omega+\varpi, \vp) \nonumber\\
&- 2 \left[\frac{\bar \Gamma^{(0,2)}_{\kappa,\parallel}(\omega+\varpi,   \vp)- 
\bar \Gamma^{(0,2)}_{\kappa,\parallel}(\varpi,   \vp)}
{\omega}\right] \times \Re \left\{\left[\frac{\bar \Gamma^{(1,1)}_{\kappa,\parallel}(\omega+\varpi, 
  \vp)- \bar \Gamma^{(1,1)}_{\kappa,\parallel}(\varpi,   \vp)}{\omega}\right]  \bar G^{(1,1)}_{\kappa,\parallel}(\omega+\varpi,  \vp) \right\}\nonumber \\
&+ \frac 1 {2\omega^2}\Bigg[ \bar \Gamma^{(0,2)}_{\kappa,\parallel}(\omega+\varpi,  \vp) -2 
\bar \Gamma^{(0,2)}_{\kappa,\parallel}(\varpi,  \vp) +\bar \Gamma^{(0,2)}_{\kappa,\parallel}(-\omega+\varpi,  \vp)  
\Bigg] \Bigg\}  \times \tilde\partial_s \int_{\vq} \bar G^{(2,0)}_{\kappa,\parallel}(\omega,  \vq). 
\label{dtGam02}
\end{align}

The two flow  equations~\eqref{dtGam11} and~\eqref{dtGam02} for the two-point functions are thus closed in the limit of large $p$. However, they still bare a complicated  nonlinear form. It turns out that they endow a much simpler form  when expressed for the correlation functions $\bar G^{(2,0)}_\kappa$ and $\bar G^{(1,1)}_\kappa$ rather than for the vertex functions $\bar \Gamma_{\kappa}^{(0,2)}$ and  $\bar \Gamma_{\kappa}^{(1,1)}$ \cite{Canet2016,Tarpin2018}.
Indeed, using the definitions~\eqref{eq:G11} and~\eqref{eq:G20},  one can calculate the flow equations for $\bar G^{(2,0)}_\kappa$ and $\bar G^{(1,1)}_\kappa$ from ~\eqref{dtGam11} and~\eqref{dtGam02}, and one finds, using parity and after some algebra 
\begin{align}
\p_s \bar G^{(2,0)}_{\kappa,\parallel}(\varpi, \vp) &=   \dfrac{p^2}{d}
\int_{\omega} \frac{1}{2\omega^2}\Big[\bar G^{(2,0)}_{\kappa,\parallel}(\omega+\varpi,  \vp) -2 
\bar G^{(2,0)}_{\kappa,\parallel}(\varpi,  \vp) +\bar G^{(2,0)}_{\kappa,\parallel}(-\omega+\varpi,  \vp) \Big] \tilde\partial_s \int_{  \vq}  \bar G^{(2,0)}_{\kappa,\parallel}(\omega,  \vq)\label{eq:flowCapp}\\
\p_s \bar G^{(1,1)}_{\kappa,\parallel}(\varpi, \vp) &=   \dfrac{p^2}{d}
\int_{\omega} \frac{1}{2\omega^2}\Big[\bar G^{(1,1)}_{\kappa,\parallel}(\omega+\varpi,  \vp) -2 
\bar G^{(1,1)}_{\kappa,\parallel}(\varpi,  \vp) +\bar G^{(1,1)}_{\kappa,\parallel}(-\omega+\varpi,  \vp) \Big] \tilde\partial_s \int_{ \vq} \bar G^{(2,0)}_{\kappa,\parallel}(\omega, \vq)\, .
\end{align}

\end{widetext}


\begin{thebibliography}{54}%
\makeatletter
\providecommand \@ifxundefined [1]{%
 \@ifx{#1\undefined}
}%
\providecommand \@ifnum [1]{%
 \ifnum #1\expandafter \@firstoftwo
 \else \expandafter \@secondoftwo
 \fi
}%
\providecommand \@ifx [1]{%
 \ifx #1\expandafter \@firstoftwo
 \else \expandafter \@secondoftwo
 \fi
}%
\providecommand \natexlab [1]{#1}%
\providecommand \enquote  [1]{``#1''}%
\providecommand \bibnamefont  [1]{#1}%
\providecommand \bibfnamefont [1]{#1}%
\providecommand \citenamefont [1]{#1}%
\providecommand \href@noop [0]{\@secondoftwo}%
\providecommand \href [0]{\begingroup \@sanitize@url \@href}%
\providecommand \@href[1]{\@@startlink{#1}\@@href}%
\providecommand \@@href[1]{\endgroup#1\@@endlink}%
\providecommand \@sanitize@url [0]{\catcode `\\12\catcode `\$12\catcode
  `\&12\catcode `\#12\catcode `\^12\catcode `\_12\catcode `\%12\relax}%
\providecommand \@@startlink[1]{}%
\providecommand \@@endlink[0]{}%
\providecommand \url  [0]{\begingroup\@sanitize@url \@url }%
\providecommand \@url [1]{\endgroup\@href {#1}{\urlprefix }}%
\providecommand \urlprefix  [0]{URL }%
\providecommand \Eprint [0]{\href }%
\providecommand \doibase [0]{https://doi.org/}%
\providecommand \selectlanguage [0]{\@gobble}%
\providecommand \bibinfo  [0]{\@secondoftwo}%
\providecommand \bibfield  [0]{\@secondoftwo}%
\providecommand \translation [1]{[#1]}%
\providecommand \BibitemOpen [0]{}%
\providecommand \bibitemStop [0]{}%
\providecommand \bibitemNoStop [0]{.\EOS\space}%
\providecommand \EOS [0]{\spacefactor3000\relax}%
\providecommand \BibitemShut  [1]{\csname bibitem#1\endcsname}%
\let\auto@bib@innerbib\@empty
\bibitem [{\citenamefont {Kardar}\ \emph {et~al.}(1986)\citenamefont {Kardar},
  \citenamefont {Parisi},\ and\ \citenamefont {Zhang}}]{Kardar86}%
  \BibitemOpen
  \bibfield  {author} {\bibinfo {author} {\bibfnamefont {M.}~\bibnamefont
  {Kardar}}, \bibinfo {author} {\bibfnamefont {G.}~\bibnamefont {Parisi}},\
  and\ \bibinfo {author} {\bibfnamefont {Y.-C.}\ \bibnamefont {Zhang}},\
  }\bibfield  {title} {\bibinfo {title} {Dynamic scaling of growing
  interfaces},\ }\href {https://doi.org/10.1103/PhysRevLett.56.889} {\bibfield
  {journal} {\bibinfo  {journal} {Phys. Rev. Lett.}\ }\textbf {\bibinfo
  {volume} {56}},\ \bibinfo {pages} {889} (\bibinfo {year} {1986})}\BibitemShut
  {NoStop}%
\bibitem [{\citenamefont {Wei}\ \emph {et~al.}(2022)\citenamefont {Wei},
  \citenamefont {Rubio-Abadal}, \citenamefont {Ye}, \citenamefont {Machado},
  \citenamefont {Kemp}, \citenamefont {Srakaew}, \citenamefont {Hollerith},
  \citenamefont {Rui}, \citenamefont {Gopalakrishnan}, \citenamefont {Yao},
  \citenamefont {Bloch},\ and\ \citenamefont {Zeiher}}]{Bloch2022}%
  \BibitemOpen
  \bibfield  {author} {\bibinfo {author} {\bibfnamefont {D.}~\bibnamefont
  {Wei}}, \bibinfo {author} {\bibfnamefont {A.}~\bibnamefont {Rubio-Abadal}},
  \bibinfo {author} {\bibfnamefont {B.}~\bibnamefont {Ye}}, \bibinfo {author}
  {\bibfnamefont {F.}~\bibnamefont {Machado}}, \bibinfo {author} {\bibfnamefont
  {J.}~\bibnamefont {Kemp}}, \bibinfo {author} {\bibfnamefont {K.}~\bibnamefont
  {Srakaew}}, \bibinfo {author} {\bibfnamefont {S.}~\bibnamefont {Hollerith}},
  \bibinfo {author} {\bibfnamefont {J.}~\bibnamefont {Rui}}, \bibinfo {author}
  {\bibfnamefont {S.}~\bibnamefont {Gopalakrishnan}}, \bibinfo {author}
  {\bibfnamefont {N.~Y.}\ \bibnamefont {Yao}}, \bibinfo {author} {\bibfnamefont
  {I.}~\bibnamefont {Bloch}},\ and\ \bibinfo {author} {\bibfnamefont
  {J.}~\bibnamefont {Zeiher}},\ }\bibfield  {title} {\bibinfo {title} {Quantum
  gas microscopy of {K}ardar-{P}arisi-{Z}hang superdiffusion},\ }\href
  {https://doi.org/10.1126/science.abk2397} {\bibfield  {journal} {\bibinfo
  {journal} {Science}\ }\textbf {\bibinfo {volume} {376}},\ \bibinfo {pages}
  {716} (\bibinfo {year} {2022})}\BibitemShut {NoStop}%
\bibitem [{\citenamefont {Scheie}\ \emph {et~al.}(2022)\citenamefont {Scheie},
  \citenamefont {Sherman}, \citenamefont {Dupont}, \citenamefont {Nagler},
  \citenamefont {Stone}, \citenamefont {Granroth}, \citenamefont {Moore},\ and\
  \citenamefont {Tennant}}]{Tennant2022}%
  \BibitemOpen
  \bibfield  {author} {\bibinfo {author} {\bibfnamefont {A.}~\bibnamefont
  {Scheie}}, \bibinfo {author} {\bibfnamefont {N.~E.}\ \bibnamefont {Sherman}},
  \bibinfo {author} {\bibfnamefont {M.}~\bibnamefont {Dupont}}, \bibinfo
  {author} {\bibfnamefont {S.~E.}\ \bibnamefont {Nagler}}, \bibinfo {author}
  {\bibfnamefont {M.~B.}\ \bibnamefont {Stone}}, \bibinfo {author}
  {\bibfnamefont {G.~E.}\ \bibnamefont {Granroth}}, \bibinfo {author}
  {\bibfnamefont {J.~E.}\ \bibnamefont {Moore}},\ and\ \bibinfo {author}
  {\bibfnamefont {D.~A.}\ \bibnamefont {Tennant}},\ }\bibfield  {title}
  {\bibinfo {title} {Detection of {K}ardar–{P}arisi–{Z}hang hydrodynamics
  in a quantum {H}eisenberg spin-1/2 chain},\ }\href
  {https://doi.org/10.1038/s41567-021-01191-6} {\bibfield  {journal} {\bibinfo
  {journal} {Nature Physics}\ }\textbf {\bibinfo {volume} {17}},\ \bibinfo
  {pages} {726} (\bibinfo {year} {2022})}\BibitemShut {NoStop}%
\bibitem [{\citenamefont {Fontaine}\ \emph {et~al.}(2022)\citenamefont
  {Fontaine}, \citenamefont {Squizzato}, \citenamefont {Baboux}, \citenamefont
  {Amelio}, \citenamefont {Lema{\^\i}tre}, \citenamefont {Morassi},
  \citenamefont {Sagnes}, \citenamefont {Le~Gratiet}, \citenamefont {Harouri},
  \citenamefont {Wouters}, \citenamefont {Carusotto}, \citenamefont {Amo},
  \citenamefont {Richard}, \citenamefont {Minguzzi}, \citenamefont {Canet},
  \citenamefont {Ravets},\ and\ \citenamefont {Bloch}}]{Fontaine2022Nat}%
  \BibitemOpen
  \bibfield  {author} {\bibinfo {author} {\bibfnamefont {Q.}~\bibnamefont
  {Fontaine}}, \bibinfo {author} {\bibfnamefont {D.}~\bibnamefont {Squizzato}},
  \bibinfo {author} {\bibfnamefont {F.}~\bibnamefont {Baboux}}, \bibinfo
  {author} {\bibfnamefont {I.}~\bibnamefont {Amelio}}, \bibinfo {author}
  {\bibfnamefont {A.}~\bibnamefont {Lema{\^\i}tre}}, \bibinfo {author}
  {\bibfnamefont {M.}~\bibnamefont {Morassi}}, \bibinfo {author} {\bibfnamefont
  {I.}~\bibnamefont {Sagnes}}, \bibinfo {author} {\bibfnamefont
  {L.}~\bibnamefont {Le~Gratiet}}, \bibinfo {author} {\bibfnamefont
  {A.}~\bibnamefont {Harouri}}, \bibinfo {author} {\bibfnamefont
  {M.}~\bibnamefont {Wouters}}, \bibinfo {author} {\bibfnamefont
  {I.}~\bibnamefont {Carusotto}}, \bibinfo {author} {\bibfnamefont
  {A.}~\bibnamefont {Amo}}, \bibinfo {author} {\bibfnamefont {M.}~\bibnamefont
  {Richard}}, \bibinfo {author} {\bibfnamefont {A.}~\bibnamefont {Minguzzi}},
  \bibinfo {author} {\bibfnamefont {L.}~\bibnamefont {Canet}}, \bibinfo
  {author} {\bibfnamefont {S.}~\bibnamefont {Ravets}},\ and\ \bibinfo {author}
  {\bibfnamefont {J.}~\bibnamefont {Bloch}},\ }\bibfield  {title} {\bibinfo
  {title} {Kardar--{P}arisi--{Z}hang universality in a one-dimensional
  polariton condensate},\ }\href {https://doi.org/10.1038/s41586-022-05001-8}
  {\bibfield  {journal} {\bibinfo  {journal} {Nature}\ }\textbf {\bibinfo
  {volume} {608}},\ \bibinfo {pages} {687} (\bibinfo {year}
  {2022})}\BibitemShut {NoStop}%
\bibitem [{\citenamefont {Nahum}\ \emph {et~al.}(2017)\citenamefont {Nahum},
  \citenamefont {Ruhman}, \citenamefont {Vijay},\ and\ \citenamefont
  {Haah}}]{Nahum2017}%
  \BibitemOpen
  \bibfield  {author} {\bibinfo {author} {\bibfnamefont {A.}~\bibnamefont
  {Nahum}}, \bibinfo {author} {\bibfnamefont {J.}~\bibnamefont {Ruhman}},
  \bibinfo {author} {\bibfnamefont {S.}~\bibnamefont {Vijay}},\ and\ \bibinfo
  {author} {\bibfnamefont {J.}~\bibnamefont {Haah}},\ }\bibfield  {title}
  {\bibinfo {title} {Quantum entanglement growth under random unitary
  dynamics},\ }\href {https://doi.org/10.1103/PhysRevX.7.031016} {\bibfield
  {journal} {\bibinfo  {journal} {Phys. Rev. X}\ }\textbf {\bibinfo {volume}
  {7}},\ \bibinfo {pages} {031016} (\bibinfo {year} {2017})}\BibitemShut
  {NoStop}%
\bibitem [{\citenamefont {Mu}\ \emph {et~al.}(2024)\citenamefont {Mu},
  \citenamefont {Gong},\ and\ \citenamefont {Lemari\'e}}]{Mu2024}%
  \BibitemOpen
  \bibfield  {author} {\bibinfo {author} {\bibfnamefont {S.}~\bibnamefont
  {Mu}}, \bibinfo {author} {\bibfnamefont {J.}~\bibnamefont {Gong}},\ and\
  \bibinfo {author} {\bibfnamefont {G.}~\bibnamefont {Lemari\'e}},\ }\bibfield
  {title} {\bibinfo {title} {Kardar-parisi-zhang physics in the density
  fluctuations of localized two-dimensional wave packets},\ }\href
  {https://doi.org/10.1103/PhysRevLett.132.046301} {\bibfield  {journal}
  {\bibinfo  {journal} {Phys. Rev. Lett.}\ }\textbf {\bibinfo {volume} {132}},\
  \bibinfo {pages} {046301} (\bibinfo {year} {2024})}\BibitemShut {NoStop}%
\bibitem [{\citenamefont {Halpin-Healy}\ and\ \citenamefont
  {Zhang}(1995)}]{Halpin-Healy95}%
  \BibitemOpen
  \bibfield  {author} {\bibinfo {author} {\bibfnamefont {T.}~\bibnamefont
  {Halpin-Healy}}\ and\ \bibinfo {author} {\bibfnamefont {Y.-C.}\ \bibnamefont
  {Zhang}},\ }\bibfield  {title} {\bibinfo {title} {Kinetic roughening
  phenomena, stochastic growth, directed polymers and all that. {A}spects of
  multidisciplinary statistical mechanics},\ }\href {https://doi.org/DOI:
  10.1016/0370-1573(94)00087-J} {\bibfield  {journal} {\bibinfo  {journal}
  {Phys. Rep.}\ }\textbf {\bibinfo {volume} {254}},\ \bibinfo {pages} {215 }
  (\bibinfo {year} {1995})}\BibitemShut {NoStop}%
\bibitem [{\citenamefont {Krug}(1997)}]{Krug97}%
  \BibitemOpen
  \bibfield  {author} {\bibinfo {author} {\bibfnamefont {J.}~\bibnamefont
  {Krug}},\ }\bibfield  {title} {\bibinfo {title} {Origins of scale invariance
  in growth processes},\ }\href {https://doi.org/10.1080/00018739700101498}
  {\bibfield  {journal} {\bibinfo  {journal} {Adv. Phys.}\ }\textbf {\bibinfo
  {volume} {46}},\ \bibinfo {pages} {139 } (\bibinfo {year}
  {1997})}\BibitemShut {NoStop}%
\bibitem [{\citenamefont {Burgers}(1948)}]{Burgers48}%
  \BibitemOpen
  \bibfield  {author} {\bibinfo {author} {\bibfnamefont {J.}~\bibnamefont
  {Burgers}},\ }\bibfield  {title} {\bibinfo {title} {A mathematical model
  illustrating the theory of turbulence}\ }(\bibinfo  {publisher} {Elsevier},\
  \bibinfo {year} {1948})\ pp.\ \bibinfo {pages} {171--199}\BibitemShut
  {NoStop}%
\bibitem [{\citenamefont {Kardar}(1987)}]{Kardar87}%
  \BibitemOpen
  \bibfield  {author} {\bibinfo {author} {\bibfnamefont {M.}~\bibnamefont
  {Kardar}},\ }\bibfield  {title} {\bibinfo {title} {Replica {B}ethe ansatz
  studies of two-dimensional interfaces with quenched random impurities},\
  }\href {https://doi.org/10.1016/0550-3213(87)90203-3} {\bibfield  {journal}
  {\bibinfo  {journal} {Nucl. Phys. B}\ }\textbf {\bibinfo {volume} {290}},\
  \bibinfo {pages} {582 } (\bibinfo {year} {1987})}\BibitemShut {NoStop}%
\bibitem [{\citenamefont {Forster}\ \emph {et~al.}(1976)\citenamefont
  {Forster}, \citenamefont {Nelson},\ and\ \citenamefont
  {Stephen}}]{Forster76}%
  \BibitemOpen
  \bibfield  {author} {\bibinfo {author} {\bibfnamefont {D.}~\bibnamefont
  {Forster}}, \bibinfo {author} {\bibfnamefont {D.~R.}\ \bibnamefont
  {Nelson}},\ and\ \bibinfo {author} {\bibfnamefont {M.~J.}\ \bibnamefont
  {Stephen}},\ }\bibfield  {title} {\bibinfo {title} {Long-time tails and the
  large-eddy behavior of a randomly stirred fluid},\ }\href
  {https://doi.org/10.1103/PhysRevLett.36.867} {\bibfield  {journal} {\bibinfo
  {journal} {Phys. Rev. Lett.}\ }\textbf {\bibinfo {volume} {36}},\ \bibinfo
  {pages} {867} (\bibinfo {year} {1976})}\BibitemShut {NoStop}%
\bibitem [{\citenamefont {Forster}\ \emph {et~al.}(1977)\citenamefont
  {Forster}, \citenamefont {Nelson},\ and\ \citenamefont
  {Stephen}}]{Forster77}%
  \BibitemOpen
  \bibfield  {author} {\bibinfo {author} {\bibfnamefont {D.}~\bibnamefont
  {Forster}}, \bibinfo {author} {\bibfnamefont {D.~R.}\ \bibnamefont
  {Nelson}},\ and\ \bibinfo {author} {\bibfnamefont {M.~J.}\ \bibnamefont
  {Stephen}},\ }\bibfield  {title} {\bibinfo {title} {Large-distance and
  long-time properties of a randomly stirred fluid},\ }\href
  {https://doi.org/10.1103/PhysRevA.16.732} {\bibfield  {journal} {\bibinfo
  {journal} {Phys. Rev. A}\ }\textbf {\bibinfo {volume} {16}},\ \bibinfo
  {pages} {732} (\bibinfo {year} {1977})}\BibitemShut {NoStop}%
\bibitem [{\citenamefont {Edwards}\ and\ \citenamefont
  {Wilkinson}(1982)}]{Edwards82}%
  \BibitemOpen
  \bibfield  {author} {\bibinfo {author} {\bibfnamefont {S.~F.}\ \bibnamefont
  {Edwards}}\ and\ \bibinfo {author} {\bibfnamefont {D.~R.}\ \bibnamefont
  {Wilkinson}},\ }\bibfield  {title} {\bibinfo {title} {{The Surface Statistics
  of a Granular Aggregate}},\ }\href {https://doi.org/10.1098/rspa.1982.0056}
  {\bibfield  {journal} {\bibinfo  {journal} {Proc. R. Soc. Lon. A}\ }\textbf
  {\bibinfo {volume} {381}},\ \bibinfo {pages} {17} (\bibinfo {year}
  {1982})}\BibitemShut {NoStop}%
\bibitem [{\citenamefont {Verma}(2000)}]{Verma2000}%
  \BibitemOpen
  \bibfield  {author} {\bibinfo {author} {\bibfnamefont {M.~K.}\ \bibnamefont
  {Verma}},\ }\bibfield  {title} {\bibinfo {title} {Intermittency exponents and
  energy spectrum of the {B}urgers and {KPZ} equations with correlated noise},\
  }\href {https://doi.org/http://dx.doi.org/10.1016/S0378-4371(99)00544-0}
  {\bibfield  {journal} {\bibinfo  {journal} {Physica A}\ }\textbf {\bibinfo
  {volume} {277}},\ \bibinfo {pages} {359 } (\bibinfo {year}
  {2000})}\BibitemShut {NoStop}%
\bibitem [{\citenamefont {Bec}\ and\ \citenamefont {Khanin}(2007)}]{Bec2007}%
  \BibitemOpen
  \bibfield  {author} {\bibinfo {author} {\bibfnamefont {J.}~\bibnamefont
  {Bec}}\ and\ \bibinfo {author} {\bibfnamefont {K.}~\bibnamefont {Khanin}},\
  }\bibfield  {title} {\bibinfo {title} {Burgers turbulence},\ }\href
  {https://doi.org/https://doi.org/10.1016/j.physrep.2007.04.002} {\bibfield
  {journal} {\bibinfo  {journal} {Physics Reports}\ }\textbf {\bibinfo {volume}
  {447}},\ \bibinfo {pages} {1} (\bibinfo {year} {2007})}\BibitemShut {NoStop}%
\bibitem [{\citenamefont {Bouchaud}\ \emph {et~al.}(1995)\citenamefont
  {Bouchaud}, \citenamefont {M\'ezard},\ and\ \citenamefont
  {Parisi}}]{Bouchaud1995}%
  \BibitemOpen
  \bibfield  {author} {\bibinfo {author} {\bibfnamefont {J.~P.}\ \bibnamefont
  {Bouchaud}}, \bibinfo {author} {\bibfnamefont {M.}~\bibnamefont {M\'ezard}},\
  and\ \bibinfo {author} {\bibfnamefont {G.}~\bibnamefont {Parisi}},\
  }\bibfield  {title} {\bibinfo {title} {Scaling and intermittency in burgers
  turbulence},\ }\href {https://doi.org/10.1103/PhysRevE.52.3656} {\bibfield
  {journal} {\bibinfo  {journal} {Phys. Rev. E}\ }\textbf {\bibinfo {volume}
  {52}},\ \bibinfo {pages} {3656} (\bibinfo {year} {1995})}\BibitemShut
  {NoStop}%
\bibitem [{\citenamefont {Cartes}\ \emph {et~al.}(2022)\citenamefont {Cartes},
  \citenamefont {Tirapegui}, \citenamefont {Pandit},\ and\ \citenamefont
  {Brachet}}]{Brachet2022}%
  \BibitemOpen
  \bibfield  {author} {\bibinfo {author} {\bibfnamefont {C.}~\bibnamefont
  {Cartes}}, \bibinfo {author} {\bibfnamefont {E.}~\bibnamefont {Tirapegui}},
  \bibinfo {author} {\bibfnamefont {R.}~\bibnamefont {Pandit}},\ and\ \bibinfo
  {author} {\bibfnamefont {M.}~\bibnamefont {Brachet}},\ }\bibfield  {title}
  {\bibinfo {title} {The {G}alerkin-truncated {B}urgers equation: crossover
  from inviscid-thermalized to {K}ardar–{P}arisi–{Z}hang scaling},\ }\href
  {https://doi.org/10.1098/rsta.2021.0090} {\bibfield  {journal} {\bibinfo
  {journal} {Philosophical Transactions of the Royal Society A: Mathematical,
  Physical and Engineering Sciences}\ }\textbf {\bibinfo {volume} {380}},\
  \bibinfo {pages} {20210090} (\bibinfo {year} {2022})}\BibitemShut {NoStop}%
\bibitem [{\citenamefont {Majda}\ and\ \citenamefont
  {Timofeyev}(2000)}]{Majda2000}%
  \BibitemOpen
  \bibfield  {author} {\bibinfo {author} {\bibfnamefont {A.~J.}\ \bibnamefont
  {Majda}}\ and\ \bibinfo {author} {\bibfnamefont {I.}~\bibnamefont
  {Timofeyev}},\ }\bibfield  {title} {\bibinfo {title} {Remarkable statistical
  behavior for truncated {B}urgers–{H}opf dynamics},\ }\href
  {https://doi.org/10.1073/pnas.230433997} {\bibfield  {journal} {\bibinfo
  {journal} {Proceedings of the National Academy of Sciences}\ }\textbf
  {\bibinfo {volume} {97}},\ \bibinfo {pages} {12413} (\bibinfo {year}
  {2000})}\BibitemShut {NoStop}%
\bibitem [{\citenamefont {Rodr\'{\i}guez-Fern\'andez}\ \emph
  {et~al.}(2022)\citenamefont {Rodr\'{\i}guez-Fern\'andez}, \citenamefont
  {Santalla}, \citenamefont {Castro},\ and\ \citenamefont
  {Cuerno}}]{Rodriguez2022}%
  \BibitemOpen
  \bibfield  {author} {\bibinfo {author} {\bibfnamefont {E.}~\bibnamefont
  {Rodr\'{\i}guez-Fern\'andez}}, \bibinfo {author} {\bibfnamefont {S.~N.}\
  \bibnamefont {Santalla}}, \bibinfo {author} {\bibfnamefont {M.}~\bibnamefont
  {Castro}},\ and\ \bibinfo {author} {\bibfnamefont {R.}~\bibnamefont
  {Cuerno}},\ }\bibfield  {title} {\bibinfo {title} {Anomalous ballistic
  scaling in the tensionless or inviscid {K}ardar-{P}arisi-{Z}hang equation},\
  }\href {https://doi.org/10.1103/PhysRevE.106.024802} {\bibfield  {journal}
  {\bibinfo  {journal} {Phys. Rev. E}\ }\textbf {\bibinfo {volume} {106}},\
  \bibinfo {pages} {024802} (\bibinfo {year} {2022})}\BibitemShut {NoStop}%
\bibitem [{\citenamefont {Fujimoto}\ \emph {et~al.}(2020)\citenamefont
  {Fujimoto}, \citenamefont {Hamazaki},\ and\ \citenamefont
  {Kawaguchi}}]{Fujimoto2020}%
  \BibitemOpen
  \bibfield  {author} {\bibinfo {author} {\bibfnamefont {K.}~\bibnamefont
  {Fujimoto}}, \bibinfo {author} {\bibfnamefont {R.}~\bibnamefont {Hamazaki}},\
  and\ \bibinfo {author} {\bibfnamefont {Y.}~\bibnamefont {Kawaguchi}},\
  }\bibfield  {title} {\bibinfo {title} {{F}amily-{V}icsek scaling of roughness
  growth in a strongly interacting bose gas},\ }\href
  {https://doi.org/10.1103/PhysRevLett.124.210604} {\bibfield  {journal}
  {\bibinfo  {journal} {Phys. Rev. Lett.}\ }\textbf {\bibinfo {volume} {124}},\
  \bibinfo {pages} {210604} (\bibinfo {year} {2020})}\BibitemShut {NoStop}%
\bibitem [{\citenamefont {Fontaine}\ \emph {et~al.}(2023)\citenamefont
  {Fontaine}, \citenamefont {Vercesi}, \citenamefont {Brachet},\ and\
  \citenamefont {Canet}}]{Fontaine2023InvBurgers}%
  \BibitemOpen
  \bibfield  {author} {\bibinfo {author} {\bibfnamefont {C.}~\bibnamefont
  {Fontaine}}, \bibinfo {author} {\bibfnamefont {F.}~\bibnamefont {Vercesi}},
  \bibinfo {author} {\bibfnamefont {M.}~\bibnamefont {Brachet}},\ and\ \bibinfo
  {author} {\bibfnamefont {L.}~\bibnamefont {Canet}},\ }\bibfield  {title}
  {\bibinfo {title} {Unpredicted scaling of the one-dimensional
  kardar-parisi-zhang equation},\ }\href
  {https://doi.org/10.1103/PhysRevLett.131.247101} {\bibfield  {journal}
  {\bibinfo  {journal} {Phys. Rev. Lett.}\ }\textbf {\bibinfo {volume} {131}},\
  \bibinfo {pages} {247101} (\bibinfo {year} {2023})}\BibitemShut {NoStop}%
\bibitem [{\citenamefont {Canet}\ \emph {et~al.}(2017)\citenamefont {Canet},
  \citenamefont {Rossetto}, \citenamefont {Wschebor},\ and\ \citenamefont
  {Balarac}}]{Canet2017}%
  \BibitemOpen
  \bibfield  {author} {\bibinfo {author} {\bibfnamefont {L.}~\bibnamefont
  {Canet}}, \bibinfo {author} {\bibfnamefont {V.}~\bibnamefont {Rossetto}},
  \bibinfo {author} {\bibfnamefont {N.}~\bibnamefont {Wschebor}},\ and\
  \bibinfo {author} {\bibfnamefont {G.}~\bibnamefont {Balarac}},\ }\bibfield
  {title} {\bibinfo {title} {Spatiotemporal velocity-velocity correlation
  function in fully developed turbulence},\ }\href
  {https://doi.org/10.1103/PhysRevE.95.023107} {\bibfield  {journal} {\bibinfo
  {journal} {Phys. Rev. E}\ }\textbf {\bibinfo {volume} {95}},\ \bibinfo
  {pages} {023107} (\bibinfo {year} {2017})}\BibitemShut {NoStop}%
\bibitem [{\citenamefont {Tarpin}\ \emph {et~al.}(2018)\citenamefont {Tarpin},
  \citenamefont {Canet},\ and\ \citenamefont {Wschebor}}]{Tarpin2018}%
  \BibitemOpen
  \bibfield  {author} {\bibinfo {author} {\bibfnamefont {M.}~\bibnamefont
  {Tarpin}}, \bibinfo {author} {\bibfnamefont {L.}~\bibnamefont {Canet}},\ and\
  \bibinfo {author} {\bibfnamefont {N.}~\bibnamefont {Wschebor}},\ }\bibfield
  {title} {\bibinfo {title} {Breaking of scale invariance in the time
  dependence of correlation functions in isotropic and homogeneous
  turbulence},\ }\href {https://doi.org/10.1063/1.5020022} {\bibfield
  {journal} {\bibinfo  {journal} {Physics of Fluids}\ }\textbf {\bibinfo
  {volume} {30}},\ \bibinfo {pages} {055102} (\bibinfo {year}
  {2018})}\BibitemShut {NoStop}%
\bibitem [{\citenamefont {Martin}\ \emph {et~al.}(1973)\citenamefont {Martin},
  \citenamefont {Siggia},\ and\ \citenamefont {Rose}}]{Martin73}%
  \BibitemOpen
  \bibfield  {author} {\bibinfo {author} {\bibfnamefont {P.~C.}\ \bibnamefont
  {Martin}}, \bibinfo {author} {\bibfnamefont {E.~D.}\ \bibnamefont {Siggia}},\
  and\ \bibinfo {author} {\bibfnamefont {H.~A.}\ \bibnamefont {Rose}},\
  }\bibfield  {title} {\bibinfo {title} {Statistical dynamics of classical
  systems},\ }\href {https://doi.org/10.1103/PhysRevA.8.423} {\bibfield
  {journal} {\bibinfo  {journal} {Phys. Rev. A}\ }\textbf {\bibinfo {volume}
  {8}},\ \bibinfo {pages} {423} (\bibinfo {year} {1973})}\BibitemShut {NoStop}%
\bibitem [{\citenamefont {Janssen}(1976)}]{Janssen76}%
  \BibitemOpen
  \bibfield  {author} {\bibinfo {author} {\bibfnamefont {H.-K.}\ \bibnamefont
  {Janssen}},\ }\bibfield  {title} {\bibinfo {title} {On a {L}agrangean for
  classical field dynamics and {R}enormalization {G}roup calculations of
  dynamical critical properties},\ }\href {https://doi.org/10.1007/BF01316547}
  {\bibfield  {journal} {\bibinfo  {journal} {Z. Phys. B}\ }\textbf {\bibinfo
  {volume} {23}},\ \bibinfo {pages} {377} (\bibinfo {year} {1976})}\BibitemShut
  {NoStop}%
\bibitem [{\citenamefont {de~Dominicis}(1976)}]{Dominicis76}%
  \BibitemOpen
  \bibfield  {author} {\bibinfo {author} {\bibfnamefont {C.}~\bibnamefont
  {de~Dominicis}},\ }\bibfield  {title} {\bibinfo {title} {Techniques de
  renormalisation de la th\'eorie des champs et dynamique des ph\'enom\`enes
  critiques},\ }\href {https://doi.org/10.1051/jphyscol:1976138} {\bibfield
  {journal} {\bibinfo  {journal} {J. Phys. (Paris) Colloq.}\ }\textbf {\bibinfo
  {volume} {37}},\ \bibinfo {pages} {247} (\bibinfo {year} {1976})}\BibitemShut
  {NoStop}%
\bibitem [{\citenamefont {Frey}\ and\ \citenamefont {T\"auber}(1994)}]{Frey94}%
  \BibitemOpen
  \bibfield  {author} {\bibinfo {author} {\bibfnamefont {E.}~\bibnamefont
  {Frey}}\ and\ \bibinfo {author} {\bibfnamefont {U.~C.}\ \bibnamefont
  {T\"auber}},\ }\bibfield  {title} {\bibinfo {title} {Two-loop
  {R}enormalization-{G}roup analysis of the
  {B}urgers--{K}ardar-{P}arisi-{Z}hang equation},\ }\href
  {https://doi.org/10.1103/PhysRevE.50.1024} {\bibfield  {journal} {\bibinfo
  {journal} {Phys. Rev. E}\ }\textbf {\bibinfo {volume} {50}},\ \bibinfo
  {pages} {1024} (\bibinfo {year} {1994})}\BibitemShut {NoStop}%
\bibitem [{\citenamefont {Canet}(2005)}]{Canet2005b}%
  \BibitemOpen
  \bibfield  {author} {\bibinfo {author} {\bibfnamefont {L.}~\bibnamefont
  {Canet}},\ }\bibfield  {title} {\bibinfo {title} {Strong-coupling fixed point
  of the {K}ardar-{P}arisi-{Z}hang equation},\ }\href
  {http://arxiv.org/abs/cond-mat/0509541} {\bibfield  {journal} {\bibinfo
  {journal} {arXiv:cond-mat/0509541}\ } (\bibinfo {year} {2005})},\ \Eprint
  {https://arxiv.org/abs/arXiv:cond-mat/0509541} {arXiv:cond-mat/0509541}
  \BibitemShut {NoStop}%
\bibitem [{\citenamefont {Canet}\ \emph {et~al.}(2016)\citenamefont {Canet},
  \citenamefont {Delamotte},\ and\ \citenamefont {Wschebor}}]{Canet2016}%
  \BibitemOpen
  \bibfield  {author} {\bibinfo {author} {\bibfnamefont {L.}~\bibnamefont
  {Canet}}, \bibinfo {author} {\bibfnamefont {B.}~\bibnamefont {Delamotte}},\
  and\ \bibinfo {author} {\bibfnamefont {N.}~\bibnamefont {Wschebor}},\
  }\bibfield  {title} {\bibinfo {title} {Fully developed isotropic turbulence:
  Nonperturbative renormalization group formalism and fixed-point solution},\
  }\href {https://doi.org/10.1103/PhysRevE.93.063101} {\bibfield  {journal}
  {\bibinfo  {journal} {Phys. Rev. E}\ }\textbf {\bibinfo {volume} {93}},\
  \bibinfo {pages} {063101} (\bibinfo {year} {2016})}\BibitemShut {NoStop}%
\bibitem [{\citenamefont {Canet}(2022)}]{Canet2022}%
  \BibitemOpen
  \bibfield  {author} {\bibinfo {author} {\bibfnamefont {L.}~\bibnamefont
  {Canet}},\ }\bibfield  {title} {\bibinfo {title} {Functional renormalisation
  group for turbulence},\ }\href {https://doi.org/10.1017/jfm.2022.808}
  {\bibfield  {journal} {\bibinfo  {journal} {Journal of Fluid Mechanics}\
  }\textbf {\bibinfo {volume} {950}},\ \bibinfo {pages} {P1} (\bibinfo {year}
  {2022})}\BibitemShut {NoStop}%
\bibitem [{\citenamefont {Canet}\ \emph {et~al.}(2015)\citenamefont {Canet},
  \citenamefont {Delamotte},\ and\ \citenamefont {Wschebor}}]{Canet2015}%
  \BibitemOpen
  \bibfield  {author} {\bibinfo {author} {\bibfnamefont {L.}~\bibnamefont
  {Canet}}, \bibinfo {author} {\bibfnamefont {B.}~\bibnamefont {Delamotte}},\
  and\ \bibinfo {author} {\bibfnamefont {N.}~\bibnamefont {Wschebor}},\
  }\bibfield  {title} {\bibinfo {title} {Fully developed isotropic turbulence:
  Symmetries and exact identities},\ }\href
  {https://doi.org/10.1103/PhysRevE.91.053004} {\bibfield  {journal} {\bibinfo
  {journal} {Phys. Rev. E}\ }\textbf {\bibinfo {volume} {91}},\ \bibinfo
  {pages} {053004} (\bibinfo {year} {2015})}\BibitemShut {NoStop}%
\bibitem [{\citenamefont {Canet}\ \emph
  {et~al.}(2011{\natexlab{a}})\citenamefont {Canet}, \citenamefont {Chat\'e},\
  and\ \citenamefont {Delamotte}}]{Canet2011heq}%
  \BibitemOpen
  \bibfield  {author} {\bibinfo {author} {\bibfnamefont {L.}~\bibnamefont
  {Canet}}, \bibinfo {author} {\bibfnamefont {H.}~\bibnamefont {Chat\'e}},\
  and\ \bibinfo {author} {\bibfnamefont {B.}~\bibnamefont {Delamotte}},\
  }\bibfield  {title} {\bibinfo {title} {General framework of the
  non-perturbative {R}enormalization {G}roup for non-equilibrium steady
  states},\ }\href {https://doi.org/10.1088/1751-8113/44/49/495001} {\bibfield
  {journal} {\bibinfo  {journal} {J. Phys. A}\ }\textbf {\bibinfo {volume}
  {44}},\ \bibinfo {pages} {495001} (\bibinfo {year}
  {2011}{\natexlab{a}})}\BibitemShut {NoStop}%
\bibitem [{\citenamefont {Berges}\ \emph {et~al.}(2002)\citenamefont {Berges},
  \citenamefont {Tetradis},\ and\ \citenamefont {Wetterich}}]{Berges2002}%
  \BibitemOpen
  \bibfield  {author} {\bibinfo {author} {\bibfnamefont {J.}~\bibnamefont
  {Berges}}, \bibinfo {author} {\bibfnamefont {N.}~\bibnamefont {Tetradis}},\
  and\ \bibinfo {author} {\bibfnamefont {C.}~\bibnamefont {Wetterich}},\
  }\bibfield  {title} {\bibinfo {title} {Non-perturbative {R}enormalization
  flow in quantum field theory and statistical physics},\ }\href
  {https://doi.org/DOI:10.1016/S0370-1573(01)00098-9} {\bibfield  {journal}
  {\bibinfo  {journal} {Phys. Rep.}\ }\textbf {\bibinfo {volume} {363}},\
  \bibinfo {pages} {223 } (\bibinfo {year} {2002})}\BibitemShut {NoStop}%
\bibitem [{\citenamefont {Kopietz}\ \emph {et~al.}(2010)\citenamefont
  {Kopietz}, \citenamefont {Bartosch},\ and\ \citenamefont
  {Sch\"utz}}]{Kopietz2010}%
  \BibitemOpen
  \bibfield  {author} {\bibinfo {author} {\bibfnamefont {P.}~\bibnamefont
  {Kopietz}}, \bibinfo {author} {\bibfnamefont {L.}~\bibnamefont {Bartosch}},\
  and\ \bibinfo {author} {\bibfnamefont {F.}~\bibnamefont {Sch\"utz}},\ }\href
  {https://doi.org/https://doi.org/10.1007/978-3-642-05094-7} {\emph {\bibinfo
  {title} {Introduction to the Functional Renormalization Group}}},\ \bibinfo
  {series} {Lecture Notes in Physics}, Vol.\ \bibinfo {volume} {798}\ (\bibinfo
   {publisher} {Springer},\ \bibinfo {address} {Berlin},\ \bibinfo {year}
  {2010})\BibitemShut {NoStop}%
\bibitem [{\citenamefont {Delamotte}(2012)}]{Delamotte2012}%
  \BibitemOpen
  \bibfield  {author} {\bibinfo {author} {\bibfnamefont {B.}~\bibnamefont
  {Delamotte}},\ }\bibfield  {title} {\bibinfo {title} {{An Introduction to the
  nonperturbative renormalization group}},\ }\href
  {https://doi.org/10.1007/978-3-642-27320-9_2} {\bibfield  {journal} {\bibinfo
   {journal} {Lect. Notes Phys.}\ }\textbf {\bibinfo {volume} {852}},\ \bibinfo
  {pages} {49} (\bibinfo {year} {2012})}\BibitemShut {NoStop}%
\bibitem [{\citenamefont {Dupuis}\ \emph {et~al.}(2021)\citenamefont {Dupuis},
  \citenamefont {Canet}, \citenamefont {Eichhorn}, \citenamefont {Metzner},
  \citenamefont {Pawlowski}, \citenamefont {Tissier},\ and\ \citenamefont
  {Wschebor}}]{Dupuis2021}%
  \BibitemOpen
  \bibfield  {author} {\bibinfo {author} {\bibfnamefont {N.}~\bibnamefont
  {Dupuis}}, \bibinfo {author} {\bibfnamefont {L.}~\bibnamefont {Canet}},
  \bibinfo {author} {\bibfnamefont {A.}~\bibnamefont {Eichhorn}}, \bibinfo
  {author} {\bibfnamefont {W.}~\bibnamefont {Metzner}}, \bibinfo {author}
  {\bibfnamefont {J.}~\bibnamefont {Pawlowski}}, \bibinfo {author}
  {\bibfnamefont {M.}~\bibnamefont {Tissier}},\ and\ \bibinfo {author}
  {\bibfnamefont {N.}~\bibnamefont {Wschebor}},\ }\bibfield  {title} {\bibinfo
  {title} {The nonperturbative functional renormalization group and its
  applications},\ }\href
  {https://doi.org/https://doi.org/10.1016/j.physrep.2021.01.001} {\bibfield
  {journal} {\bibinfo  {journal} {Physics Reports}\ }\textbf {\bibinfo {volume}
  {910}},\ \bibinfo {pages} {1} (\bibinfo {year} {2021})},\ \bibinfo {note}
  {the nonperturbative functional renormalization group and its
  applications}\BibitemShut {NoStop}%
\bibitem [{\citenamefont {Wetterich}(1993)}]{Wetterich93}%
  \BibitemOpen
  \bibfield  {author} {\bibinfo {author} {\bibfnamefont {C.}~\bibnamefont
  {Wetterich}},\ }\bibfield  {title} {\bibinfo {title} {Exact evolution
  equation for the effective potential},\ }\href {https://doi.org/DOI:
  10.1016/0370-2693(93)90726-X} {\bibfield  {journal} {\bibinfo  {journal}
  {Phys. Lett. B}\ }\textbf {\bibinfo {volume} {301}},\ \bibinfo {pages} {90 }
  (\bibinfo {year} {1993})}\BibitemShut {NoStop}%
\bibitem [{\citenamefont {Ellwanger}(1994)}]{Ellwanger94}%
  \BibitemOpen
  \bibfield  {author} {\bibinfo {author} {\bibfnamefont {U.}~\bibnamefont
  {Ellwanger}},\ }\bibfield  {title} {\bibinfo {title} {Flow equations and
  {BRS} invariance for {Y}ang-{M}ills theories},\ }\href
  {https://doi.org/https://doi.org/10.1016/0370-2693(94)90365-4} {\bibfield
  {journal} {\bibinfo  {journal} {Physics Letters B}\ }\textbf {\bibinfo
  {volume} {335}},\ \bibinfo {pages} {364} (\bibinfo {year}
  {1994})}\BibitemShut {NoStop}%
\bibitem [{\citenamefont {Morris}(1994)}]{Morris94}%
  \BibitemOpen
  \bibfield  {author} {\bibinfo {author} {\bibfnamefont {T.~R.}\ \bibnamefont
  {Morris}},\ }\bibfield  {title} {\bibinfo {title} {The exact renormalization
  group and approximate solutions},\ }\href
  {https://doi.org/10.1142/S0217751X94000972} {\bibfield  {journal} {\bibinfo
  {journal} {International Journal of Modern Physics A}\ }\textbf {\bibinfo
  {volume} {09}},\ \bibinfo {pages} {2411} (\bibinfo {year}
  {1994})}\BibitemShut {NoStop}%
\bibitem [{\citenamefont {Benitez}\ and\ \citenamefont
  {Wschebor}(2013)}]{Benitez13}%
  \BibitemOpen
  \bibfield  {author} {\bibinfo {author} {\bibfnamefont {F.}~\bibnamefont
  {Benitez}}\ and\ \bibinfo {author} {\bibfnamefont {N.}~\bibnamefont
  {Wschebor}},\ }\bibfield  {title} {\bibinfo {title} {Branching and
  annihilating random walks: Exact results at low branching rate},\ }\href
  {https://doi.org/10.1103/PhysRevE.87.052132} {\bibfield  {journal} {\bibinfo
  {journal} {Phys. Rev. E}\ }\textbf {\bibinfo {volume} {87}},\ \bibinfo
  {pages} {052132} (\bibinfo {year} {2013})}\BibitemShut {NoStop}%
\bibitem [{\citenamefont {Canet}\ \emph {et~al.}(2010)\citenamefont {Canet},
  \citenamefont {Chat\'e}, \citenamefont {Delamotte},\ and\ \citenamefont
  {Wschebor}}]{Canet2010}%
  \BibitemOpen
  \bibfield  {author} {\bibinfo {author} {\bibfnamefont {L.}~\bibnamefont
  {Canet}}, \bibinfo {author} {\bibfnamefont {H.}~\bibnamefont {Chat\'e}},
  \bibinfo {author} {\bibfnamefont {B.}~\bibnamefont {Delamotte}},\ and\
  \bibinfo {author} {\bibfnamefont {N.}~\bibnamefont {Wschebor}},\ }\bibfield
  {title} {\bibinfo {title} {Nonperturbative renormalization group for the
  {K}ardar-{P}arisi-{Z}hang equation},\ }\href
  {https://doi.org/10.1103/PhysRevLett.104.150601} {\bibfield  {journal}
  {\bibinfo  {journal} {Phys. Rev. Lett.}\ }\textbf {\bibinfo {volume} {104}},\
  \bibinfo {pages} {150601} (\bibinfo {year} {2010})}\BibitemShut {NoStop}%
\bibitem [{\citenamefont {Canet}\ \emph
  {et~al.}(2011{\natexlab{b}})\citenamefont {Canet}, \citenamefont {Chat\'e},
  \citenamefont {Delamotte},\ and\ \citenamefont {Wschebor}}]{Canet2011kpz}%
  \BibitemOpen
  \bibfield  {author} {\bibinfo {author} {\bibfnamefont {L.}~\bibnamefont
  {Canet}}, \bibinfo {author} {\bibfnamefont {H.}~\bibnamefont {Chat\'e}},
  \bibinfo {author} {\bibfnamefont {B.}~\bibnamefont {Delamotte}},\ and\
  \bibinfo {author} {\bibfnamefont {N.}~\bibnamefont {Wschebor}},\ }\bibfield
  {title} {\bibinfo {title} {Nonperturbative {R}enormalization {G}roup for the
  {K}ardar-{P}arisi-{Z}hang equation: General framework and first
  applications},\ }\href {https://doi.org/10.1103/PhysRevE.84.061128}
  {\bibfield  {journal} {\bibinfo  {journal} {Phys. Rev. E}\ }\textbf {\bibinfo
  {volume} {84}},\ \bibinfo {pages} {061128} (\bibinfo {year}
  {2011}{\natexlab{b}})}\BibitemShut {NoStop}%
\bibitem [{\citenamefont {Kloss}\ \emph {et~al.}(2012)\citenamefont {Kloss},
  \citenamefont {Canet},\ and\ \citenamefont {Wschebor}}]{Kloss2012}%
  \BibitemOpen
  \bibfield  {author} {\bibinfo {author} {\bibfnamefont {T.}~\bibnamefont
  {Kloss}}, \bibinfo {author} {\bibfnamefont {L.}~\bibnamefont {Canet}},\ and\
  \bibinfo {author} {\bibfnamefont {N.}~\bibnamefont {Wschebor}},\ }\bibfield
  {title} {\bibinfo {title} {Nonperturbative {R}enormalization {G}roup for the
  stationary {K}ardar-{P}arisi-{Z}hang equation: Scaling functions and
  amplitude ratios in 1+1, 2+1, and 3+1 dimensions},\ }\href
  {https://doi.org/10.1103/PhysRevE.86.051124} {\bibfield  {journal} {\bibinfo
  {journal} {Phys. Rev. E}\ }\textbf {\bibinfo {volume} {86}},\ \bibinfo
  {pages} {051124} (\bibinfo {year} {2012})}\BibitemShut {NoStop}%
\bibitem [{\citenamefont {Wiese}(1998)}]{Wiese98}%
  \BibitemOpen
  \bibfield  {author} {\bibinfo {author} {\bibfnamefont {K.~J.}\ \bibnamefont
  {Wiese}},\ }\bibfield  {title} {\bibinfo {title} {On the perturbation
  expansion of the {{KPZ}} equation},\ }\href
  {https://doi.org/10.1023/B:JOSS.0000026730.76868.c4} {\bibfield  {journal}
  {\bibinfo  {journal} {J. Stat. Phys.}\ }\textbf {\bibinfo {volume} {93}},\
  \bibinfo {pages} {143} (\bibinfo {year} {1998})}\BibitemShut {NoStop}%
\bibitem [{Note1()}]{Note1}%
  \BibitemOpen
  \bibinfo {note} {Again, the statement that the system is already at the KPZ
  fixed point for $g_\Lambda =g_*$ only holds within the simplest approximation
  \protect \eqref {eq:anzLPA} where $\Gamma _\kappa $ is restricted to the form
  of the bare action. In general, the fixed point effective action is more
  complicated, and $g_*$ is defined as the initial value $g_\Lambda $ which
  minimizes the RG time to reach the IR fixed point.}\BibitemShut {Stop}%
\bibitem [{\citenamefont {Blaizot}\ \emph {et~al.}(2006)\citenamefont
  {Blaizot}, \citenamefont {M\'endez-Galain},\ and\ \citenamefont
  {Wschebor}}]{Blaizot2006}%
  \BibitemOpen
  \bibfield  {author} {\bibinfo {author} {\bibfnamefont {J.-P.}\ \bibnamefont
  {Blaizot}}, \bibinfo {author} {\bibfnamefont {R.}~\bibnamefont
  {M\'endez-Galain}},\ and\ \bibinfo {author} {\bibfnamefont {N.}~\bibnamefont
  {Wschebor}},\ }\bibfield  {title} {\bibinfo {title} {A new method to solve
  the non-perturbative {R}enormalization {G}roup equations},\ }\href
  {https://doi.org/DOI: 10.1016/j.physletb.2005.10.086} {\bibfield  {journal}
  {\bibinfo  {journal} {Phys. Lett. B}\ }\textbf {\bibinfo {volume} {632}},\
  \bibinfo {pages} {571 } (\bibinfo {year} {2006})}\BibitemShut {NoStop}%
\bibitem [{\citenamefont {Blaizot}\ \emph {et~al.}(2007)\citenamefont
  {Blaizot}, \citenamefont {M\'endez-Galain},\ and\ \citenamefont
  {Wschebor}}]{Blaizot2007}%
  \BibitemOpen
  \bibfield  {author} {\bibinfo {author} {\bibfnamefont {J.-P.}\ \bibnamefont
  {Blaizot}}, \bibinfo {author} {\bibfnamefont {R.}~\bibnamefont
  {M\'endez-Galain}},\ and\ \bibinfo {author} {\bibfnamefont {N.}~\bibnamefont
  {Wschebor}},\ }\bibfield  {title} {\bibinfo {title} {Non-perturbative
  {R}enormalization {G}roup calculation of the scalar self-energy},\ }\href
  {https://doi.org/10.1140/epjb/e2007-00223-3} {\bibfield  {journal} {\bibinfo
  {journal} {Eur. Phys. J. B}\ }\textbf {\bibinfo {volume} {58}},\ \bibinfo
  {pages} {297} (\bibinfo {year} {2007})}\BibitemShut {NoStop}%
\bibitem [{\citenamefont {Benitez}\ \emph {et~al.}(2012)\citenamefont
  {Benitez}, \citenamefont {Blaizot}, \citenamefont {Chat\'e}, \citenamefont
  {Delamotte}, \citenamefont {M\'endez-Galain},\ and\ \citenamefont
  {Wschebor}}]{Benitez2012}%
  \BibitemOpen
  \bibfield  {author} {\bibinfo {author} {\bibfnamefont {F.}~\bibnamefont
  {Benitez}}, \bibinfo {author} {\bibfnamefont {J.-P.}\ \bibnamefont
  {Blaizot}}, \bibinfo {author} {\bibfnamefont {H.}~\bibnamefont {Chat\'e}},
  \bibinfo {author} {\bibfnamefont {B.}~\bibnamefont {Delamotte}}, \bibinfo
  {author} {\bibfnamefont {R.}~\bibnamefont {M\'endez-Galain}},\ and\ \bibinfo
  {author} {\bibfnamefont {N.}~\bibnamefont {Wschebor}},\ }\bibfield  {title}
  {\bibinfo {title} {Nonperturbative {R}enormalization {G}roup preserving
  full-momentum dependence: Implementation and quantitative evaluation},\
  }\href {https://doi.org/10.1103/PhysRevE.85.026707} {\bibfield  {journal}
  {\bibinfo  {journal} {Phys. Rev. E}\ }\textbf {\bibinfo {volume} {85}},\
  \bibinfo {pages} {026707} (\bibinfo {year} {2012})}\BibitemShut {NoStop}%
\bibitem [{\citenamefont {Tarpin}\ \emph {et~al.}(2019)\citenamefont {Tarpin},
  \citenamefont {Canet}, \citenamefont {Pagani},\ and\ \citenamefont
  {Wschebor}}]{Tarpin2019}%
  \BibitemOpen
  \bibfield  {author} {\bibinfo {author} {\bibfnamefont {M.}~\bibnamefont
  {Tarpin}}, \bibinfo {author} {\bibfnamefont {L.}~\bibnamefont {Canet}},
  \bibinfo {author} {\bibfnamefont {C.}~\bibnamefont {Pagani}},\ and\ \bibinfo
  {author} {\bibfnamefont {N.}~\bibnamefont {Wschebor}},\ }\bibfield  {title}
  {\bibinfo {title} {Stationary, isotropic and homogeneous two-dimensional
  turbulence: a first non-perturbative renormalization group approach},\ }\href
  {https://doi.org/10.1088/1751-8121/aaf3f0} {\bibfield  {journal} {\bibinfo
  {journal} {Journal of Physics A: Mathematical and Theoretical}\ }\textbf
  {\bibinfo {volume} {52}},\ \bibinfo {pages} {085501} (\bibinfo {year}
  {2019})}\BibitemShut {NoStop}%
\bibitem [{\citenamefont {Pagani}\ and\ \citenamefont
  {Canet}(2021)}]{Pagani2021}%
  \BibitemOpen
  \bibfield  {author} {\bibinfo {author} {\bibfnamefont {C.}~\bibnamefont
  {Pagani}}\ and\ \bibinfo {author} {\bibfnamefont {L.}~\bibnamefont {Canet}},\
  }\bibfield  {title} {\bibinfo {title} {Spatio-temporal correlation functions
  in scalar turbulence from functional renormalization group},\ }\href
  {https://doi.org/10.1063/5.0050515} {\bibfield  {journal} {\bibinfo
  {journal} {Physics of Fluids}\ }\textbf {\bibinfo {volume} {33}},\ \bibinfo
  {pages} {065109} (\bibinfo {year} {2021})}\BibitemShut {NoStop}%
\bibitem [{\citenamefont {Gorbunova}\ \emph
  {et~al.}(2021{\natexlab{a}})\citenamefont {Gorbunova}, \citenamefont
  {Balarac}, \citenamefont {Canet}, \citenamefont {Eyink},\ and\ \citenamefont
  {Rossetto}}]{Gorbunova2021}%
  \BibitemOpen
  \bibfield  {author} {\bibinfo {author} {\bibfnamefont {A.}~\bibnamefont
  {Gorbunova}}, \bibinfo {author} {\bibfnamefont {G.}~\bibnamefont {Balarac}},
  \bibinfo {author} {\bibfnamefont {L.}~\bibnamefont {Canet}}, \bibinfo
  {author} {\bibfnamefont {G.}~\bibnamefont {Eyink}},\ and\ \bibinfo {author}
  {\bibfnamefont {V.}~\bibnamefont {Rossetto}},\ }\bibfield  {title} {\bibinfo
  {title} {Spatio-temporal correlations in three-dimensional homogeneous and
  isotropic turbulence},\ }\href {https://doi.org/10.1063/5.0046677} {\bibfield
   {journal} {\bibinfo  {journal} {Physics of Fluids}\ }\textbf {\bibinfo
  {volume} {33}},\ \bibinfo {pages} {045114} (\bibinfo {year}
  {2021}{\natexlab{a}})}\BibitemShut {NoStop}%
\bibitem [{\citenamefont {Gorbunova}\ \emph
  {et~al.}(2021{\natexlab{b}})\citenamefont {Gorbunova}, \citenamefont
  {Pagani}, \citenamefont {Balarac}, \citenamefont {Canet},\ and\ \citenamefont
  {Rossetto}}]{Gorbunova2021scalar}%
  \BibitemOpen
  \bibfield  {author} {\bibinfo {author} {\bibfnamefont {A.}~\bibnamefont
  {Gorbunova}}, \bibinfo {author} {\bibfnamefont {C.}~\bibnamefont {Pagani}},
  \bibinfo {author} {\bibfnamefont {G.}~\bibnamefont {Balarac}}, \bibinfo
  {author} {\bibfnamefont {L.}~\bibnamefont {Canet}},\ and\ \bibinfo {author}
  {\bibfnamefont {V.}~\bibnamefont {Rossetto}},\ }\bibfield  {title} {\bibinfo
  {title} {Eulerian spatiotemporal correlations in passive scalar turbulence},\
  }\href {https://doi.org/10.1103/PhysRevFluids.6.124606} {\bibfield  {journal}
  {\bibinfo  {journal} {Phys. Rev. Fluids}\ }\textbf {\bibinfo {volume} {6}},\
  \bibinfo {pages} {124606} (\bibinfo {year} {2021}{\natexlab{b}})}\BibitemShut
  {NoStop}%
\bibitem [{\citenamefont {Kelling}\ and\ \citenamefont
  {\'Odor}(2011)}]{Kelling11}%
  \BibitemOpen
  \bibfield  {author} {\bibinfo {author} {\bibfnamefont {J.}~\bibnamefont
  {Kelling}}\ and\ \bibinfo {author} {\bibfnamefont {G.}~\bibnamefont
  {\'Odor}},\ }\bibfield  {title} {\bibinfo {title} {Extremely large-scale
  simulation of a {K}ardar-{P}arisi-{Z}hang model using graphics cards},\
  }\href {https://doi.org/10.1103/PhysRevE.84.061150} {\bibfield  {journal}
  {\bibinfo  {journal} {Phys. Rev. E}\ }\textbf {\bibinfo {volume} {84}},\
  \bibinfo {pages} {061150} (\bibinfo {year} {2011})}\BibitemShut {NoStop}%
\bibitem [{\citenamefont {Oliveira}\ \emph {et~al.}(2013)\citenamefont
  {Oliveira}, \citenamefont {Alves},\ and\ \citenamefont
  {Ferreira}}]{Oliveira13}%
  \BibitemOpen
  \bibfield  {author} {\bibinfo {author} {\bibfnamefont {T.~J.}\ \bibnamefont
  {Oliveira}}, \bibinfo {author} {\bibfnamefont {S.~G.}\ \bibnamefont
  {Alves}},\ and\ \bibinfo {author} {\bibfnamefont {S.~C.}\ \bibnamefont
  {Ferreira}},\ }\bibfield  {title} {\bibinfo {title}
  {{K}ardar-{P}arisi-{Z}hang universality class in ($2+1$) dimensions:
  Universal geometry-dependent distributions and finite-time corrections},\
  }\href {https://doi.org/10.1103/PhysRevE.87.040102} {\bibfield  {journal}
  {\bibinfo  {journal} {Phys. Rev. E}\ }\textbf {\bibinfo {volume} {87}},\
  \bibinfo {pages} {040102} (\bibinfo {year} {2013})}\BibitemShut {NoStop}%
\end{thebibliography}
%

\end{document}